\documentclass[twocolumn,preprintnumbers,aps,pra,toc]{revtex5}

\usepackage{graphicx}
\usepackage{dcolumn}
\usepackage{bm}
\usepackage{amssymb,amsmath}
\usepackage{dsfont}
\usepackage[usenames]{color}
\usepackage{multirow}

\def\bea{\begin{eqnarray}}
\def\eea{\end{eqnarray}}
\def\ben{\begin{equation}}
\def\een{\end{equation}}
\def\benu{\begin{enumerate}}
\def\enu{\end{enumerate}}

\def\n{n}

\def\sss{\scriptscriptstyle\rm}

\def\l{^\lambda}

\def\1var{(\bx_1...\bx\N)}

\def\half{\frac{1}{2}}

\def\br{{\bf r}}

\def\bx{{\br t}}

\def\bj{{\bf j}}

\def\x{_{\sss X}}

\def\s{_{\sss S}}
\def\xc{_{\sss XC}}

\def\N{_{\sss N}}
\def\H{_{\sss H}}

\def\ext{_{\rm ext}}

\def\adia{^{\rm adia}}

\def\unif{^{\rm unif}}

\def\ee{_{\rm ee}}

\def\ALDA{^{\rm ALDA}}

\def\sph_int{ {\int d^3 r}}

\def\PRL{Phys. Rev. Letts.\ }
\def\JCP{J. Chem. Phys.\ }

\input epsf
\input psfig.sty
\input rotate

\def\bei{\begin{itemize}}
\def\eei{\end{itemize}}
\def\bx{{\br t}}
\def\bbbone{\mathds{1}}

\def\sig{_{\sigma}}
\def\sigp{_{\sigma'}}
\def\dsig{_{\sigma\sigma'}}

\def\dssum{\sum_{\sigma\sigma'}}
\def\ssump{\sum_{\sigma'}}
\def\sS{_{{\sss S} \sigma}}
\def\extS{_{\rm ext\sigma}}
\def\xcS{_{{\sss XC}\sigma}}
\def\extSp{_{\rm ext\sigma'}}
\def\sSp{_{{\sss S} \sigma'}}

\begin{document} 

\title{Excited states from time-dependent density functional theory}
\author{Peter Elliott}
\affiliation{Department of Physics and Astronomy, University of California, Irvine, CA 92697, USA}
\author{Kieron Burke}
\affiliation{Department of Chemistry, University of California, Irvine, CA 92697, USA}
\author{Filipp Furche}
\affiliation{Institut f\"{u}r Physikalische Chemie, Universit\"{a}t Karlsruhe, Kaiserstra\ss e 12, 76128 Karlsruhe, Germany}
\date{\today}


\begin{abstract}
Time-dependent density functional theory (TDDFT) is presently enjoying 
enormous popularity in quantum chemistry, as a useful tool for 
extracting electronic excited state energies.
This article explains what TDDFT is, and how it differs from
ground-state DFT.  We show the basic formalism, and illustrate
with simple examples.  We discuss its implementation and possible
sources of error.  We discuss many of the major successes and
challenges of the theory, including weak fields, strong fields,
continuum states, double excitations, charge transfer, high harmonic
generation, multiphoton ionization, electronic quantum control,
van der Waals interactions, transport through single molecules,
currents, quantum defects, and, elastic electron-atom scattering.
\end{abstract}
\maketitle
\tableofcontents
\newpage
\section{Introduction}
\label{s:intro}

Ground-state density functional theory
\cite{HK64,KS65,DG90} has become the method of choice
for calculating ground-state properties of large molecules, because it replaces
the interacting many-electron problem with an effective
single-particle problem that can be solved much more quickly.
It is based on rigorous theorems\cite{HK64,KS65,L82}
and a hierarchy of increasingly accurate approximations,
such as the local density approximation (LDA), generalized
gradient approximations (GGA's)\cite{B88,LYP88,PBE96}, and hybrids
of exact exchange with GGA\cite{Bb93}.
For example, a recent ground-state calculation\cite{FP06} for 
crambin ($\mbox{C}_{203}\mbox{H}_{317}\mbox{N}_{55}\mbox{O}_{64}\mbox{S}_6$), a small protein, using 
TURBOMOLE\cite{TURBO} on a $1.5$ GHZ HP itanium workstation took just $6$h$52$m, extraordinarily 
fast for 2528 electrons. But, formally, ground-state density functional theory predicts only ground-state properties, 
not electronic excitations.

On the other hand, time-dependent density functional theory (TDDFT)\cite{MG04,FB05,MBAG01,BG98,GDP96} applies
the same philosophy to time-dependent problems.
We replace the complicated many-body time-dependent Schr\"odinger
equation by a set of time-dependent single-particle equations whose
orbitals yield the same time-dependent density.
We can do this because the Runge-Gross theorem\cite{RG84}
proves that, for a given initial wavefunction,
particle statistics and interaction, a given time-dependent
density can arise from at most one time-dependent external
potential.   This means that the time-dependent potential (and all other properties)
is a functional of the time-dependent density.

Armed with a formal theorem, we can then
define time-dependent Kohn-Sham (TDKS) equations
that describe non-interacting electrons that evolve in a time-dependent Kohn-Sham
potential,
but produce the same density as that of the interacting
system of interest.  Thus, just as in the ground-state case,
the demanding interacting time-dependent Schr\"odinger
equation is replaced by a much simpler set of equations to
propagate.The price of this enormous simplification is that the
exchange-correlation piece of the Kohn-Sham potential has to be
approximated.

The most common time-dependent perturbation is a long-wavelength
electric field, oscillating with frequency $\omega$.  In the usual
situation, this field is a weak perturbation on the molecule, and
one can perform a linear response analysis.  From this, we can
extract the optical absorption spectrum of the molecule due to
electronic excitations.  Thus linear response TDDFT predicts the
transition frequencies to electronic excited states and many other
properties.  This has been the primary use of TDDFT so far,
with lots of applications to large molecules.

\begin{figure}
  \centering
  \includegraphics[scale=0.8]{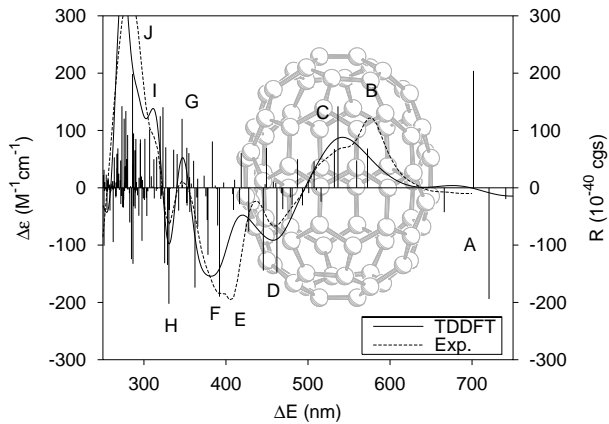}
  \caption{TDDFT calculation and experiment for the electronic CD
    spectrum of fullerene ($^{\text{f}}\!A$)-C$_{76}$. 
    TDDFT calculations were performed with the BP86 functional
    and an augmented SVP basis set \protect\cite{Furche02d}. The
    RI-$J$ approximation together with TZVP auxiliary basis sets
    \protect\cite{Eichkorn97a} was used. Experimental data (in
    CH$_2$Cl$_2$) are from \protect\cite{Goto98a}.} 
  \label{f:c76}
\end{figure}

Figure \ref{f:c76} compares TDDFT and experiment for the electronic CD
spectrum of the chiral fullerene C$_{76}$. A total of 240 optically
allowed transitions were required to simulate the spectrum. The
accuracy is clearly good enough to assign the absolute configuration
of $C_{76}$. TDDFT calculations of this size typically take less than
a day on low-end personal computers.
 
A random walk through some recent papers using TDDFT gives
some feeling for the breadth of applications.  Most are in the
linear response regime.
In inorganic chemistry, the optical response of many transition metal
complexes\cite{CHHJ06,YSFW06,VZKH06,SHNK05,SVCL05,GRPD05,NBNA05,LRC06,FFSC06,HB06,PCLK05,VSHL04,CHBG04,SVR04,AFS04,JAZ03} has been calculated,
and even some X-ray absorption\cite{PM06,FSRD04}.
In organic chemistry, heterocycles\cite{YKYS06,MAGS06,FDBL06,LC05,FGS05,BDHC05} among 
others\cite{WHDG05,RB06,KS06} have been examined. Other examples include the response of 
thiouracil\cite{SL04},
s-tetrazine\cite{OKH04},
and annulated porphyrins\cite{RNHM03}.
We also see TDDFT's use in studying various fullerenes\cite{THCR06,MWMR04,MRT06,EWFK03,EFWK05,MRRV03}.
In biochemistry, TDDFT is finding many 
uses\cite{QRS06,NS05,TRR05,TK05,WHSK05,RWXZ06,GS06,KA06,SJZ06,SBMS06,KJDM05}.  DNA bases are under examination, and an 
overview of their study may be found in Ref. \cite{VFMR06}. 
In photobiology, potential energy curves for the trans-cis
photo-isomerization of protonated Schiff base of retinal\cite{TI04}
have been calculated.  Large calculations for green and blue
fluorescent proteins have also been performed\cite{MLVC03,LMCR05}.
Doing photochemistry with TDDFT\cite{CDIC06}, 
properties of chromophores\cite{DG06,CHBA05,KIBM05,WXRY05,TSFD06} and 
dyes\cite{JPWP06,CBC02,SJZ05,JPSF06,CGZ06,JBP06,LLCA06} have been computed.
For these and other systems, there is great interest in
charge-transfer excitations\cite{CKNG06,BZD06,GML06,BSH04,JC04,DH04,RF04,T03,JL03},
but (as we
later discuss) intermolecular charge transfer is a demanding problem for TDDFT.

Another major area of application is clusters, large and small, covalent and
metallic, and everything 
inbetween\cite{RH06,IOJ05,JAS05,GCC05,SFF05,ZLZ06,PTC06,LS06,LCWS06,GCC06,BB06,BGLP06,DHSM04,GGJI04,BS04,PRS04,RSCG04,DRGM04,BNW04,MCRA06},
including Met-Cars\cite{MCRP04}.
Several studies include solvent
effects\cite{LAJ06,NLBW05,MWMS05,ZGBA06,CKG06,GSb06,SKB06,JPSFb06,SFMT06}, one example being
the behavior of metal ions
in explicit water\cite{BBSV04}.
TDDFT in linear response
can also be used to examine
chirality\cite{MA05,SMCF05,DGb06,AJST06}, including calculating both electric and magnetic
circular dichroism\cite{NBNA05,JAZ05,JSDA06,SMBS04,SZBA04,LHG05},
and has been applied to both helical aromatics\cite{WSTI04}
and to artemisinin complexes in solution\cite{MMME04}.  
There have also been applications in materials\cite{WAT04,GKPE04} and quantum 
dots\cite{HMW04} 
but, as discussed
below, the
optical response of bulk solids requires some non-local approximations\cite{BSVO04}.

Beyond the linear regime, there is also
growing interest in second- and third-order
response\cite{IASM04,JZ04,KMT04,MT04} in all these fields.
In particular the field of non-linear optics has been
heavily investigated\cite{GYYS06,YSQ06,MTK06}, especially the phenomenon 
of two photon absorption\cite{DNP06,YQSD05,KTMW05,DNP05,DNP06,OK06,C06,RLA06,KTAB06}. 

In fact, TDDFT yields predictions for a huge variety of phenomena, 
that can largely be classified into three groups:
(i) the non-perturbative regime, with systems in laser fields
so intense that perturbation theory fails,
(ii) the linear (and higher-order) regime, which yields the
usual optical response and electronic transitions, and
(iii) back to the ground-state, where the fluctuation-dissipation
theorem produces {\em ground-state} approximations from TDDFT
treatments of excitations.

\subsection{Overview}
\label{s:over}

This work focuses primarily on the linear response regime.
Throughout, we emphasize the difference between small
molecules (atoms, diatomics, etc.) and the larger molecules that
are of greater practical interest, where TDDFT
is often the only practical first-principles method.
We use napthalene (C$_{10}$H$_{8}$) as an example to show how the
selection of the basis set and of the XC functional affects
excitation energies and oscillator strengths computed using
TDDFT. Small molecules are somewhat exceptional because they usually
exhibit high symmetry which prevents strong mixing of the KS states due
to configuration interaction; also, basis set requirements are
often exacerbated for small systems. Naphthalene is large enough to
avoid these effects, yet reasonably accurate gas phase experiments and correlated
wavefunction calculations are still available.

We use
atomic units throughout ($e^2 = \hbar = m_e = 1$), so that 
all energies are in Hartrees (1 H$\simeq$27.2 eV$\simeq$627.5 kcal/mol)
and distances in Bohr ($\simeq$0.529 \AA) unless otherwise noted.
For brevity, we
drop comma's between arguments wherever the meaning is clear.
In DFT and TDDFT, there is a confusing wealth of acronyms and abbreviations.
Table \ref{tab:useful} is  
designed to aid the readers navigation through this maze.

\begingroup
\squeezetable
\begin{table}[h]
\caption{\label{tab:useful} Table of acronyms and abbreviations.}
\begin{tabular}{r|l}
\hline
AC		&	Asymptotically corrected \\
ALDA	&	Adiabatic LDA \\
A		&	Adiabatic  \\
B88		&	Becke GGA \\
B3LYP	&	Hybrid functional using Becke exchange and LYP correlation \\
CASPT2	&	Complete active space $2^{nd}$ order perturbation theory\\
CC		&	Coupled cluster \\
CIS 	&	Configuration-interaction singlets \\
ee		&	electron-electron \\
ext		&	external \\
EXX		&	Exact exchange \\  
GGA		&	Generalized gradient approximation \\
HK 		&	Hohenberg-Kohn \\
H		&	Hartree \\
HXC		&	Hartree plus exchange-correlation \\
KS 		& 	Kohn-Sham \\
LB94	&	van Leeuwen-Baerends asymptotically corrected functional \\
LDA		&	Local density approximation \\
LSDA	&	Local spin density approximation \\
LHF		&   Localized Hartree-Fock (accurate approximation to EXX)\\
LYP		& 	Lee-Yang-Parr correlation \\
MAE		&	Mean absolute error\\
OEP		&	Optimised effective potential\\
PBE		&	Perdew-Burke-Ernzerhof GGA \\
PBE0	&	Hybrid based on PBE \\
RG		&	Runge-Gross \\
RPA		&	Random phase approximation \\
SPA		& 	Single pole approximation \\
TDKS	&	Time-dependent Kohn-Sham \\
XC		&	Exchange-correlation \\
\hline 
\end{tabular}
\end{table}
\endgroup

The content of this review is organized as follows.  Sections \ref{s:gsrev} and  \ref{s:basic} cover the
basic formalism of the theory, that is needed to understand where it
comes from, why it works, and where it can be expected to fail.
Section \ref{s:imp} is all about details of implementation, especially basis-set
selection.  On the other hand, section \ref{s:perf} is devoted to performance, and
analyzing the sources of error in the basis-set limit.  
In section \ref{s:atoms}, we then look at a few atoms in microscopic detail:  this is
because we know the {\em exact} ground-state Kohn-Sham potential in such
cases, and so can analyze TDDFT performance in great depth.  Section \ref{s:beyond} is devoted
to the many attempts to go beyond standard functional approximations, and
especially discusses where such attempts are needed.  The last substantial
section, section \ref{s:other}, covers topics outside the usual linear response approach to 
excitations, including ground-state functionals derived from TDDFT,
challenges for strong fields, and transport through single molecules.
Section \ref{s:sum} is a summary.

\section{Ground-state review}
\label{s:gsrev}

In this section, we review ground state DFT rather quickly. For a more comprehensive review, we 
recommend \cite{FNM03}. Many of the results discussed here are referred to in later sections.

\subsection{Formalism}

Ground-state DFT is a completely different approach to solving the many-electron problem than the traditional solution of the Schr\"{o}dinger equation.
The Hohenberg-Kohn (HK) theorem\cite{HK64} of $1964$ states that for
a given non-degenerate ground-state density $\n(\br)$
of Fermions with a given interaction, 
the external potential $v\ext(\br)$ that produced it is unique (up to an additive constant).
Hence if the density is known, then $v\ext(\br)$ is known and so $\hat{H}$, the Hamiltonian, is known.
From this and the number of particles (determined by the integral of the density),
all properties of the system may be determined. In particular,
the ground-state energy of the system $E$ would be known.
This is what we mean when we say these properties
are functionals of the density, e.g., $E[\n]$.
It was later shown that this holds even for degenerate ground-states\cite{L82},
and modern DFT calculations use an analogous theorem applied to 
the spin densities, $n_\alpha(\br),n_\beta(\br)$. Where $\alpha,\beta = \pm\half$ respectfully.

The total energy for $N$ electrons consists of three parts:
the kinetic energy $T[\Psi]$,
the electron-electron interaction $V\ee[\Psi]$,
and the external potential energy $V\ext[\Psi]$, each of which is defined below: 
\bea
T[\Psi] &=& \langle\Psi|-\frac{1}{2}\sum_{i=1}^{N}\nabla_{i}^{2} |\Psi\rangle \ , \\
V\ee[\Psi] &=& \langle\Psi|\frac{1}{2}\sum_{i}^{N}\sum_{j\neq i}^{N}\frac{1}{|\br_{i} - \br_{j}|}|\Psi\rangle \ , \\
V\ext[\Psi] &=& \langle\Psi|\sum_{i}^{N}v\ext(\br_{i})|\Psi\rangle \ .
\eea
By the Rayleigh-Ritz principle:

\bea
\label{rayl}
E &=& \min_{\Psi}\langle \Psi | \hat{H} | \Psi \rangle  \\
  &=& \min_{\Psi} \left(  T[\Psi] + V\ee[\Psi] + V\ext[\Psi] \right) \nonumber \ .
\eea
If we simply rewrite the minimization as a two step process\cite{L79}:  
\ben
\label{levy}  
E = \min_{\n_\alpha,\n\beta}\left\{ \min_{\Psi\rightarrow(\n_\alpha,\n_\beta)} \left(  T[\Psi] + V\ee[\Psi] + V\ext[\Psi] \right)\right\} \nonumber ,
\een
where the inner search is over all interacting wavefunctions yielding spin densities $\n_\alpha,\n_\beta$. We may pull the 
last term out of the inner minimization:
\bea
 E &=& \min_{\n_{\alpha},\n_{\beta}} \left( F[\n_{\alpha},\n_{\beta}] + \sum\sig\int \ d^{3}r \ v\extS(\br) \ \n\sig(\br) \right)  \nonumber \\
  &=& \min_{\n_{\alpha},\n_{\beta}} \left( E[\n_{\alpha},\n_{\beta}] \right) \label{denfun} \ ,
\eea
where
\bea
F[\n_\alpha,\n_\beta] &=& \min_{\Psi\rightarrow(\n_\alpha,\n_\beta)}\left( T[\Psi] + V\ee[\Psi] \right) \\
                      &=& T[\n_{\alpha},\n_{\beta}] + V\ee[\n_{\alpha},\n_{\beta}] \ , \label{unifun}
\eea
is a universal functional independent of $v\extS(\br)$.\\

Minimizing the total energy density functional, Eq. (\ref{denfun}), for both spin densities by taking the functional derivative $\delta / \delta\n\sig$, 
and using the Euler-Lagrange multiplier technique leads to the equation:
\ben
\label{eulerT}
\frac{\delta F[\n_\alpha,\n_\beta]}{\delta\n\sig} + v\extS(\br) = \mu \ ,
\een
where $\mu$ is the chemical potential of the system.\\

Next we imagine a system of non-interacting electrons with the same spin densities. Applying the HK theorem to this non-interacting system, the potentials, $v\sS(\br)$, that give
densities $\n\sig(\br)$ as the ground-state spin densities for this system are unique.
This is the fictitious Kohn-Sham (KS) system\cite{KS65},
and the fully interacting problem is mapped to a non-interacting one
which gives the exact same density. Solving the KS equations, which
is computationally simple (at least compared to the fully interacting problem 
which becomes intractable for large particle numbers), then
yields the ground-state density. 
The KS equations are
\ben
\left( -\frac{1}{2}\nabla^{2}+v\sS(\br)\right)\phi_{j\sigma}(\br) = \epsilon_{j\sigma} \phi_{j\sigma}(\br) \ ,
\een
with spin densites 
\ben
\n\sig(\br) = \sum_{j=1}^{N\sig}|\phi_{j\sigma}(\br)|^{2} \ ,
\een
where $v_{{\sss S}\alpha}$,$v_{{\sss S}\beta}$ are the KS potentials and $N\sig$ is the number of spin $\sigma$ electrons
, ($N_\alpha + N_\beta = N$).\\

\begin{figure}[htb]
\unitlength1cm
\begin{picture}(12,14)
\put(-5,-6){\makebox(12,14){
\includegraphics{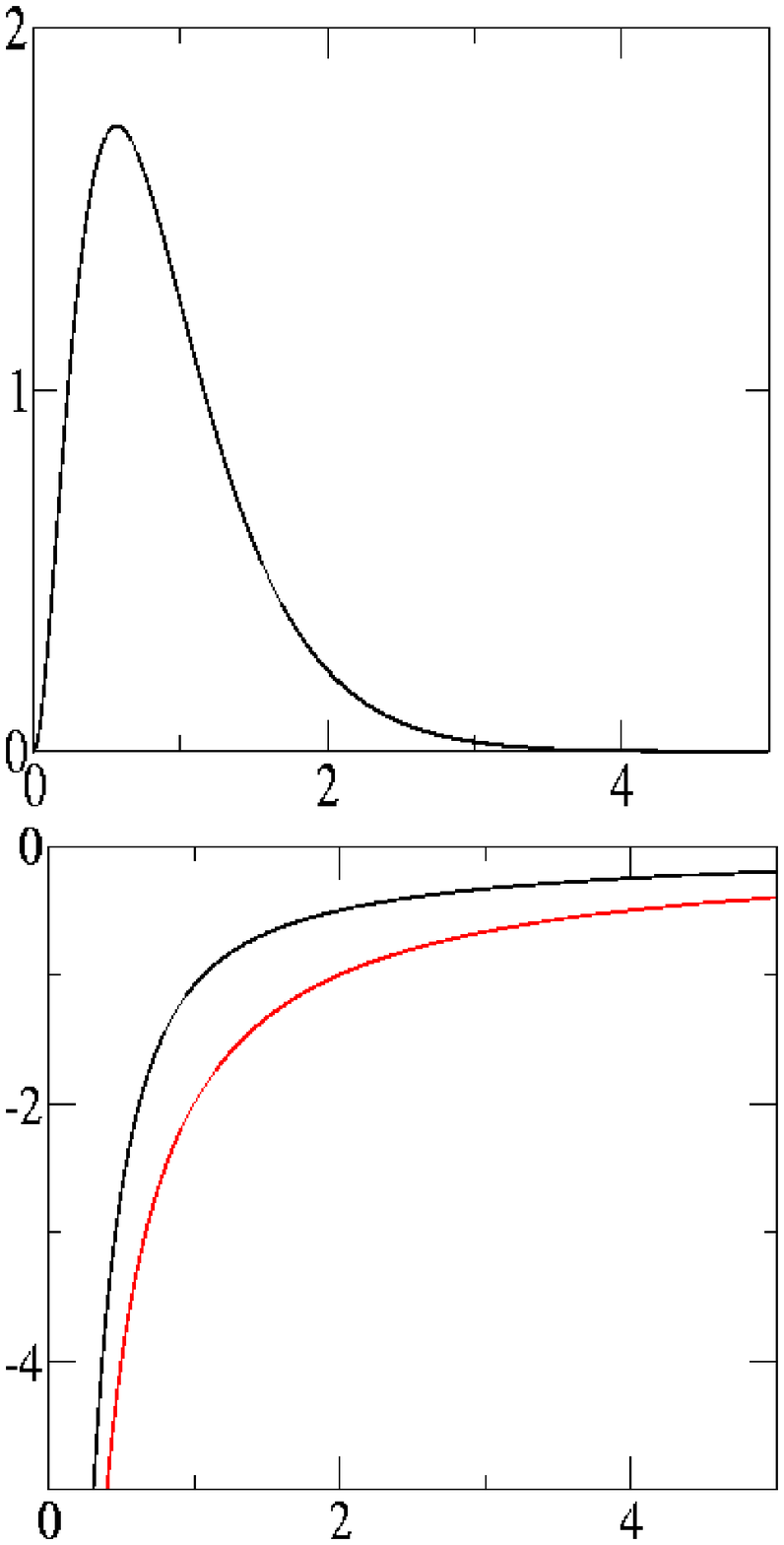}
}}
\setbox6=\hbox{\large $4\pi r^2 \n(r)$}
\put(0.6,11.3) {\makebox(0,0){\rotl 6}}
\put(4.6,7.5){\large $r$}
\setbox6=\hbox{\large $v(r)$}
\put(0.6,4.3) {\makebox(0,0){\rotl 6}}
\put(4.6,1.1){\large $r$}
\put(3.4,5.2){$v\ext(r)$}
\put(1.9,6){$v\s(r)$}
\end{picture}
\caption{Top panel -- exact radial density for the He atom found via the QMC method\cite{UG94}.  Bottom panel -- The external and KS potentials for the He atom. The KS potential is found by inverting the KS equations using the exact KS orbitals (easily found for He if exact density is known).}
\label{f:He_ext_ks}
\end{figure}

In Fig \ref{f:He_ext_ks}, we plot the exact density for the He atom from a highly accurate wavefunction calculation, and 
below we plot the {\em exact} KS potential for this system. One can see that the KS potential is very different from the external potential. This is due to the fact that the KS single effective potential for the non-interacting system must give the correct interacting electron density. Because the coulomb repulsion between the electrons shields the nucleus, and makes the charge density decay less rapidly than $e^{-4\br}$, the KS potential is shallower than $v\ext(\br)$.\\

To derive an expression for $v\sS(\br)$ we note that the Euler equation that yields the KS equations is:
\ben
\label{eulerKS}
\frac{\delta T\s[\n_\alpha,\n_\beta]}{\delta\n\sig} + v\sS(\br) = \mu \ .
\een
Here $T\s$ is the kinetic energy of the KS electrons,
\ben
T\s = \sum\sig\sum_{j=1}^{N\sig}\int~d^{3}r~\half |\nabla\phi_{j\sigma}(\br)|^{2}  \nonumber \ .
\een  
If we rewrite $F[\n_\alpha,\n_\beta]$ in terms of the KS system:
\ben
\label{unifunks}
F[\n_{\alpha},\n_{\beta}]  =  T\s[\n_\alpha,\n_\beta] + U[\n] + E\xc[\n_\alpha,\n_\beta] \ ,
\een
where $U[\n]$ is the Hartree energy, given by
\ben
U[\n] = \frac{1}{2}\dssum\int~d^{3}rd^{3}r^{\prime}~\frac{\n\sig(\br)\n\sigp(\br^{\prime})}{|\br - \br^{\prime}|} \ ,
\een
and the exchange-correlation (XC) energy is defined by Eq. (\ref{unifun}) and Eq. (\ref{unifunks}):
\ben
E\xc[\n_\alpha,\n_\beta] = T[\n_{\alpha},\n_{\beta}] - T\s[\n_\alpha,\n_\beta] + V\ee[\n_{\alpha},\n_{\beta}] - U[\n] \ .
\een
Inserting this into Eq. (\ref{eulerT}) and comparing to Eq. (\ref{eulerKS}) gives a definition of the KS potential:
\ben
v\sS(\br) = v\ext(\br) + v\H(\br) + v\xcS(\br) \ ,
\een
where the Hartree potential is the functional derivative of $U[n]$:
\ben
\label{hartpot}
v\H(\br) = \frac{\delta U[n]}{\delta\n(\br)} = \ssump\int~\ d^{3}r^{\prime}~\frac{\n\sigp(\br^{\prime})}{|\br - \br^{\prime}|} \ ,
\een
while the XC potential is given by the functional derivative of the XC energy:
\ben
v\xcS(\br) = \frac{\delta E[\n_{\alpha},\n_{\beta}]}{\delta\n\sig(\br)} \ .
\een

This then closes the relationship between the KS system and the original physical problem. Once $E\xc[\n_\alpha,\n_\beta]$ is known exactly or approximated, $v\xcS(\br)$ is determined by differentiation. The KS equations can be solved self-consistently for the spin densities and orbitals, and the total energy found by inserting these into the total energy functional $E = T\s + U + E\xc + V\ext$. Unfortunately $E\xc[\n_{\alpha},\n_{\beta}]$ is not known exactly and must be approximated. There exists a \textit{functional soup} of many
different approximations of varying accuracy and computational cost. Many of these are discussed in Section \ref{s:gsappr}.\\

In Fig \ref{f:He_ext_ks}, when the KS equation is solved with this exact potential, the HOMO level is at $-24.592~\rm$eV. This is minus the 
ionization energy for Helium. In exact DFT, Koopman's theorem, which states $I = -\epsilon_{HOMO}$, is exactly 
true\cite{PPLB82}. 
In ground-state DFT, this is the only energy level of the fictitious KS system that has an immediate physical interpretation.\\

Before leaving the ground-state review, we mention the optimized effective potential (OEP) method\cite{E03,GKKG98}. 
Here the XC functional is written as a functional of the KS orbitals (which in turn are functionals of the density). 
The exchange energy is then given by the familiar HF definition.
\ben
\label{hfxx}
E\x = -\half\sum_{i,j=1}^N \sum_{\sigma} \int d\br d\br' \
\frac{\phi_{i\sigma}^{*}(\br)\phi_{j\sigma}^{*}(\br')\phi_{j\sigma}(\br)\phi_{i\sigma}(\br')}{|\br-\br'|}
\een
However in contrast to HF, a single effective potential $v\s^{\rm {\sss XX}}(\br)$ is found via 
the chain rule. Accurate orbital dependent functionals for the correlation energy are extremely difficult to find, so often only exchange is used.
In DFT, this is called exact exchange (EXX), since exchange is usually only treated approximately. EXX gives useful features such as derivative 
discontinuities and the correct asymptotic decay of the KS potential\cite{G05}. As we will see in Section \ref{s:gspot}, these are important for TDDFT linear response.\\

\subsection{Approximate Functionals}
\label{s:gsappr}

In any ground-state DFT calculation,
we must use approximations for the functional dependence
of the XC energy on the spin densities.  There now exists a hierarchy of such
approximations.
The simplest of these is the local density approximation (LDA), where the XC energy at a point $\br^{\prime}$ is calculated 
as if it were a uniform electron gas with the spin densities $\n\sig = \n\sig(\br')$ in a constant positive background.
The exchange energy for the uniform gas can be deduced analytically, but the correlation contribution is found using a combination of 
many-body theory and highly accurate Monte Carlo simulations for the electron gas of different densities\cite{CA80,PZ81,VWN80,PW92}. \\

LDA works remarkably well, given the vast different between homogeneous electron gases and atoms or molecules. However total energies are generally
underestimated. Typically the XC energy is underestimated by about $7 \% $. When the performance of LDA is examined carefully, this comes about
via a nice (but not completely accidental) cancellation of errors between the exchange part (underestimated by about $10\%$) and correlation (overestimated by $200\%-300\%$), which
 is typically $4$ times smaller than exchange.\\

An obvious improvement to LDA would be to include information about how rapidly the density is changing via its gradient. This leads to the
generalized gradient approximation (GGA). In the original Kohn-Sham paper of $1965$, the simplest of these was suggested. The gradient expansion
approximation (GEA) is found by examining the slowly varying limit of the electron gas\cite{KS65,MB68}. However it was soon found that GEA failed
to improve on the accuracy of LDA and sometimes made things worse. It was not until the late $80$'s that accurate GGA's were constructed. The most 
popular of these are BLYP (B88\cite{B88} for exchange and LYP\cite{LYP88} for correlation) and PBE\cite{PBE96}. These generally reduce atomization errors of
 LDA by a factor of $2-5$.\\

PBE is a functional designed  to improve upon the performance of LDA without losing the features of LDA which are correct. As such it 
reduces to LDA for the uniform electron gas.
A GGA should also satisfy as  many exact conditions as possible, such as the Lieb-Oxford bound or the form of the exchange energy 
in the slowly varying limit of the electron gas.  In this regard, PBE is a 
non-empirical functional where all parameters are determined by exact conditions. Because of its 
ability to treat bulk metals, it is the functional of choice in solid-state calculations and increasingly 
so in quantum chemistry. When choosing a GGA, using PBE or not PBE should no longer be a question. (Although the crambin 
calculation of the introduction used BP86, for reasons explained in \cite{FP06}.)\\

Finally, hybrid functionals mix in some fraction of exact exchange with a GGA. This is the Hartree-Fock exchange integral, Eq. (\ref{hfxx}), evaluated
 with the KS orbitals (which are functionals of the density). Only a small fraction of
exact exchange ($20\%-25\%$) is mixed in, in order to preserve the cancellation of errors which GGA's make use of\cite{BEP97}. The most widely
 used functional in chemistry is the hybrid function B3LYP, 
which contains 3 experimentally fitted parameters\cite{B93,Bb93,LYP88} (although the parameter in B88 has recently been derived \cite{PCSB06}). Other hybrid functionals include PBE0, where $25\%$ of exact exchange is mixed
in with the PBE functional\cite{PEB96}.\\

\begin{figure}[htb]
\unitlength1cm
\begin{picture}(12,4)
\put(-6.6,-1.2){\makebox(12,3.5){
\includegraphics{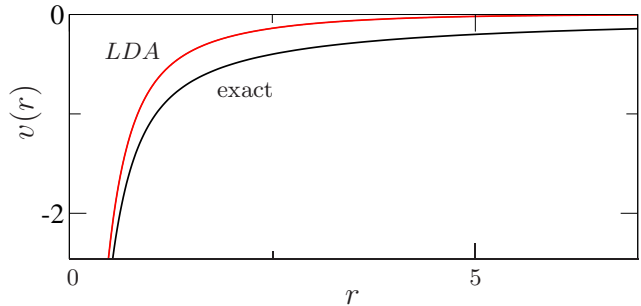}
}}
\setbox6=\hbox{\large $v(r)$}
\put(-0.05,2.5) {\makebox(0,0){\rotl 6}}
\put(4.2,-0.05){\large $r$}
\put(2.5,2.7){exact}
\put(1,3.2){$LDA$}
\put(0.53,0.55){\line(1,0){7.565}}
\put(3.24,0.55){\line(0,1){0.1}}
\put(5.92,0.55){\line(0,1){0.2}}
\put(0.5,0.2){0}
\put(5.85,0.2){5}
\end{picture}
\caption{Exact and LDA KS potentials for the He atom. While the exact potential falls off as $-1/r$, the LDA decays much too quickly. This is common for nearly all present functionals and has major consequences for TDDFT.}
\label{f:LDA_vs_exKS}
\end{figure}
A less well-known feature to users of ground state DFT is that while their favourite approximations
yield very good energies (and therefore structures, vibrations, thermochemistry, etc.)
and rather good densities, they have poorly behaved potentials, at least far from nuclei.
Figure \ref{f:LDA_vs_exKS} illustrates this for the He atom, showing the LDA potential
compared to the exact KS potential. While the potential is generally good in the region $r<2$, it decays much too fast far from the nucleus. 
The true KS potential falls off as $-1/r$ whereas LDA decays exponentially. Hence the KS eigenvalues and eigenvalues will be poor for the
 higher levels. To understand why poor potentials do not imply poor energies (and why these potentials are not as bad as they look), see Ref. \cite{BCL98}. But, as we shall see in section \ref{s:perf}, this has major consequences for TDDFT. 

Over the past decade, the technology for treating orbital-dependent functionals has developed, and such functionals 
help cure this problem\cite{E03}. This is called 
the optimized effective potential (OEP)\cite{GPGb02,PGG98,UGG95}.
The first useful orbital functional was the self-interaction corrected LDA of Perdew and Zunger\cite{PZ81}.  More generally, the OEP
method can handle any orbital-dependent functional including treating exchange exactly. Orbital-dependent functionals naturally avoid the
self-interaction error that is common in density functionals. An (almost) exact implementation of the OEP equations is 
localized Hartree-Fock (LHF)\cite{SGb01,SG02}, 
available in TURBOMOLE\cite{TURBO}.

\subsection{Basis Sets}
\label{s:fbs}

To actually solve the KS equations, the KS orbitals
$\phi_{p\sigma}(\br)$ are expanded in a \emph{finite} set of basis
functions $\chi_{\nu}(\br)$,
\begin{equation}
  \label{eq:lcao}
  \phi_{p\sigma}(\br t) = \sum_{\nu} C_{p\nu \sigma}~\chi_{\nu}(\br).
\end{equation}
The most common choice by far for the basis
functions in quantum chemistry are atom-centered contracted Cartesian
Gaussians\cite{Davidson86a}, 
\begin{equation}
  \label{eq:gaussian}
  \chi_{\nu}(\br) = \sum_{i} c_{i\nu} x^{l_x(\nu)} y^{l_y(\nu)} z^{l_z(\nu)}
  e^{-\zeta_{i\nu} (\br - {\bf R}_{\nu})^2}.
\end{equation}
$l_x(\nu)$, $l_y(\nu)$, and $l_z(\nu)$ are positive integers or zero,
and $l(\nu)=l_x(\nu)+l_y(\nu)+l_z(\nu)$ is somewhat loosely called
$l$-quantum number of $\chi_{\nu}$. ($l=0,1,2,3,\ldots$ corresponds to
$s,p,d,f,\ldots$ type Cartesian Gaussians.) The exponents $\zeta_{i\mu}$ and
the contraction coefficients $c_{i\nu}$ are optimized in atomic
calculations. Other common basis functions in use are Slater type orbitals,
plane waves, or piecewise defined functions on a numerical grid. 

The approximation of the orbitals $\phi_{p\sigma}(\br)$ by a finite
linear combination of basis functions (also called LCAO, linear
combination of atomic orbitals), Eq. (\ref{eq:lcao}), leads to a finite
number of MOs. Thus, the KS equations and all derived equations are
approximated by \emph{finite}-dimensional matrix equations. These
equations can be treated by established numerical linear and
non-linear algebra methods. When the basis set size is systematically
increased, the computed properties converge to their basis set limit.

In a finite basis set, all operators become finite matrices; the
matrix elements are integrals, e.g.,
\begin{equation}
  H_{\mu\nu \sigma}[n] = \int d^3r\, \chi_{\mu}(\br) H_{\sigma}[n]
  \chi_{\nu}(\br).
\end{equation}
The calculation and processing of such integrals is the main effort
in virtually all DFT calculations. Gaussian basis functions have the
distinct advantage that most integrals can be evaluated analytically, and that
they are spatially local. The latter implies that many integrals
vanish and need not be calculated. Whether a certain integral vanishes
or not can be decided in advance by so-called pre-screening techniques\cite{Haeser89a}.

\begin{table}
\caption{\label{tab:naphtcomb}
Single-point calculations using PBE functional for the reaction energy for naphthalene combustion using PBE/TZVP/RI geometries. Reference value computed using standard enthalpies of formation(from NIST\cite{ALS05}) using thermal and ZVPE corrections at the PBE/TZVP/RI level.}
\begin{ruledtabular}
\begin{tabular}{c | c}
Basis set & Negative reaction energy (kcal/mol) \\
\hline
SV				& 	916.8 \\
SV(P)			&	1060.0 \\
6-31G*			&	1047.1 \\
SVP				&	1108.6 \\
aug-SV(P)		&	1115.5 \\
TZVP			&	1124.5 \\
TZVPP			&	1131.2 \\
cc-pVTZ			&	1129.0 \\
aug-TZVP		&	1130.2 \\
aug-TZVP/RI		&	1130.2 \\
QZVP			&	1140.3 \\
\hline
Reference Value &	1216.3 \\
\end{tabular}
\end{ruledtabular}
\end{table}

To illustrate the effect of choosing various basis sets, we show in Fig. \ref{tab:naphtcomb}, the reaction energy for naphthalene combustion in the gas phase:
\ben
\mbox{C}_{10}\mbox{H}_8 + 12\mbox{O}_2 \rightarrow 10\mbox{C}\mbox{O}_2 + 4\mbox{H}_2\mbox{O}
\een 
The basis sets are listed in order of increasing size, and are well-know in quantum chemistry (and are described in detail in section \ref{s:imp}).
We see that hydrogen polarization functions (basis sets ending in P) are important, because C-H bonds are 
broken and O-H bonds are formed. Augmentation (aug-) with diffuse functions somewhat improves the smaller basis-set results, but is not economical in this case. Using the resolution of the identity for the Coulomb operator (RI) saves computational time, with no loss of accuracy. Reasonable results are found with SVP, but convergence improves all the way to TZVP.
We can see after the TZVPP result, the basis set error is below the functional error and the result is 
effectively converged. We have reached the stage where adding more orbitals, which increases the computational cost, is no longer going to drastically improve the result. (On the other hand, the crambin calculation mentioned in the introduction is very large and so only an SV(P) basis set could be used).\\

On first impression, comparison to the reference value indicates quite a large error, $\Delta$E$ =76$~kcal/mol. 
However, given that $48$ electron pair bonds are broken and formed, the error per carbon atom, $7.6$~kcal/mol, is typical for this functional.

\section{Time-dependent theory}
\label{s:basic}

In this section, we introduce all the basis elements of TDDFT, and how it differs from the ground-state case.

\subsection{Runge-Gross theorem}

\label{s:RG}

The analog of the Hohenberg-Kohn theorem for time-dependent
problems is the Runge-Gross theorem\cite{RG84}, which 
we state here.
Consider $N$ non-relativistic
electrons, mutually interacting via the Coulomb repulsion, in
a time-dependent external potential.  The Runge-Gross theorem
states that
the densities $\n(\br t)$ and $\n'(\br t)$ evolving from a common
initial state $\Psi_{0}=\Psi(t=0)$ under the influence of two
external potentials $v\ext(\br t)$ and $v'\ext(\br t)$ (both Taylor expandable
about the initial time $0$) are always different provided that the
potentials differ by more than a purely time-dependent (${\bf
r}$-independent) function:
\ben
\label{vdif}
\Delta v\ext(\br t) \neq c(t) \ ,
\een
where
\begin{equation}
\Delta v\ext(\br t)=v\ext(\br t)-v'\ext (\br t) \ .
\end{equation}
Thus there is a one-to-one mapping between densities and potentials,
and we say that the time-dependent potential is a functional of
the time-dependent density (and the initial state).

The theorem was proven in two distinct parts.  In the first (RGI),
one shows that the 
corresponding
current densities differ.  The current density is given by
\begin{equation} \label{jpara}
{\bf j}(\br t) = \langle \Psi(t) \vert\ \hat{{\bf j}}({\bf r})\ \vert
\Psi(t) \rangle
\een
where
\ben
\label{joper}
\hat{\bf j}({\bf r}) = \frac{1}{2i}\sum_{j=1}^{N}
\left(
\nabla_j \delta({\bf r}-{\bf r}_j)
+  \delta({\bf r}-{\bf r}_j)  \nabla_j
\right)
\end{equation}
is the current density operator.
The equation of motion for the difference of the two current densities gives\cite{RG84}:
\ben
\frac{\partial\Delta j(\bx)}{\partial t}\bigg\vert_{t=0} = -\n_{0}(\br)\nabla\Delta v\ext(\br,0)
\een
If the Taylor-expansion about $t=0$ of the difference of
the two potentials is not spatially uniform for some order, then
the Taylor-expansion of the current density difference will be non-zero
at a finite order.
This establishes that the external potential is
a functional of the  current density, $v\ext[~{\bf j},\Psi_{0}](\br,t)$.

In the second part of the theorem (RGII), continuity is used:
\begin{equation} \label{cont}
\frac{\partial  \n(\br t)}{\partial t} =
- \nabla \cdot \bj(\br t)
\end{equation}
which leads to:
\begin{equation} \label{dkdtndif}
\frac{\partial^2 \Delta \n(\br t)}{\partial t^2} 
\bigg\vert_{t=0} = \nabla \cdot \left( n_{0}({\bf r})
\nabla \Delta v\ext(\br,0) \right) 
\end{equation}
Now, suppose $\Delta v\ext(\br,0)$ is not uniform everywhere.
Might not the left-hand-side still vanish?  Apparently
not, for real systems, because it is easy to show\cite{GK90}:
\bea
\label{veciden}
\int &\!d^{3}r &  \Delta v\ext(\br,0) \nabla \cdot \left( n_{0}({\bf r})
\nabla \Delta v\ext(\br,0) \right) \nonumber\\
&=& \int d^{3}r \ \left[ \nabla\cdot \left( \Delta v\ext(\br,0)\n_{0}(\br)\nabla\Delta v\ext(\br,0)\right) \right. \nonumber \\
& & \left. \hspace{1.3cm} - \n_{0}|\nabla\Delta v\ext(\br,0) |^{2} \ \right]
\eea
Using Green's theorem, the first term on the right
vanishes for
physically realistic potentials
(i.~e.,~potentials arising from normalizable external charge densities),
because for such potentials, $\Delta v\ext(\br)$
falls off at least as $1/r$.
But the second term is definitely negative, so if $\Delta v\ext(\br,0)$ is
non-uniform, the integral must be finite, causing the
densities to differ in 2nd order in $t$.
This argument applies to each order
and the densities $\n({\bf r},t)$ and $n'({\bf r},t)$ will
become different infinitesimally later than $t$. Thus, by imposing this 
boundary conditions, we have shown that $v\ext[\n,\Psi_0](\br t)$.\\

Notes:

\begin{itemize}

\item
The difference between $\n(\bx)$ and
$\n'(\bx)$ is non-vanishing already in first order of
$\Delta v\ext(\bx)$, ensuring
the invertibility of the linear response operators of section
\ref{s:linresp}.

\item
Since the density determines the potential up to a time-dependent
constant, the wavefunction is in turn determined up to a
time-dependent phase, which cancels out of the expectation
value of any operator.

\item
We write
\ben
v\ext[\n;\Psi_0](\bx) \nonumber
\een
because it depends on both the history of the density
and the initial wavefunction.
This functional is a very complex one, much more so than the
ground-state case.   Knowledge of it implies solution of all
time-dependent Coulomb interacting problems.

\item
If we always begin in a non-degenerate
ground-state\cite{MBW02,MB01}, the initial-state dependence can be subsumed by
the Hohenberg-Kohn theorem\cite{HK64}, and  then $v\ext(\bx)$
is a functional of
$\n(\br t)$ alone:
\ben
v\ext[\n](\bx) \nonumber
\een

\item
A spin-dependent generalization exists, so that $v\ext(\br t)$ will be a functional of the spin densities $\n_{\alpha},\n_\beta$\cite{LV89}. 
\end{itemize}

\subsection{Kohn-Sham equations}
\label{s:KS}

Once we have a proof that the potential is a functional of the time-dependent density, it is simple to
write the TD Kohn-Sham (TDKS)
equations as
\ben
i \frac{\partial\phi_{j\sigma} (\br t)}{\partial t}  = 
\left( - \frac{\nabla^{2}}{2} + v\sS[n](\br t) \right)
\phi_{j\sigma}(\br t)\quad ,
\label{TDKS}
\een
whose potential is uniquely chosen (via the RG theorem) 
to reproduce the exact spin densities:
\ben 
n\sig(\br t) = \sum_{j=1}^{N\sig} | \phi_{j\sigma}(\br t)|^{2} \ ,
\een
of the interacting system.   We {\em define} the
exchange-correlation potential via
\ben
v\sS(\br t) = v\extS(\br t) + 
\int d^3r' \ \frac{\n(\br't)}{|\br-\br'|} + v\xcS (\br t) \ ,
\label{vxc}
\een
where the second term is the familiar Hartree potential.\\

Notes:

\begin{itemize}

\item
The exchange-correlation potential, $v\xcS(\bx)$,  
is in general a functional of the entire history
of the densities, $\n\sig(\bx)$, the initial interacting wavefunction
$\Psi(0)$, and the initial Kohn-Sham wavefunction, $\Phi(0)$\cite{MB01}.
But if both the KS and interacting initial wavefunctions are
non-degenerate ground-states, it becomes a simple
functional of
$n\sig(\br t)$ alone.

\item
By inverting the single doubly-occupied KS equation
for a spin-unpolarized two-electron system, it is
quite straightforward (but technically demanding) to
find the TDKS potential from an exact time-dependent density,
and has been done several times\cite{AV99,HMB02,LK05}.

\item
In practical calculations, some approximation is used
for $v\xc(\bx)$ as a functional of the density, and so
modifications of traditional TDSE schemes are needed for the propagation\cite{CMR04}.

\item
Unlike the ground-state case, there is no self-consistency,
merely forward propagation in time, of a density dependent Hamiltonian.

\item
Again, contrary to the ground-state, there is no central role
played by a single-number functional, such as the ground-state
energy.  In fact, an action was written down in the RG paper, but
extremizing it was later shown not to yield the TDKS equations\cite{L99}.

\end{itemize}

\subsection{Linear response}
\label{s:linresp}

The most common application is the response to a weak
long-wavevlength optical field, 
\ben
\delta v\ext(\br t) = -\xi\exp(i\omega t){\bf z} \ .
\een
In the general case of the response of the ground-state to an arbitrary weak external field, the system's
first-order response is characterized by the non-local susceptibility
\ben
\label{chi}
\delta \n\sig(\br t) = \ssump\int dt' \int d^3r' 
\chi\dsig[\n_0](\br,\br';t-t')\ \delta v\extSp(\br' t').
\een
This susceptibility
$\chi$ is a functional of the {\em ground-state} density, $\n_0(\br)$.
A similar equation describes the density response in the KS 
system:
\ben
\label{chis}
\delta \n\sig(\br t) = \ssump\int dt' \int d^3r' 
\chi_{\rm{\sss S}\sigma\sigma'}[\n_0](\br,\br';t-t')\ \delta v\sSp(\br' t').
\een
Here $\chi\s$ is the {\em Kohn-Sham} response function, constructed
from KS energies and orbitals:
\ben
\label{chis_l}
\chi_{{\sss S}\sigma\sigma'}(\br\br'\omega) = \delta\dsig \sum_{q}
\left\{
\frac{\Phi_{q\sigma}(\br)\ \Phi_{q\sigma^{\prime}}^{*}(\br')}
{\omega-\omega_q+i 0_+} - 
\frac{\Phi_{q\sigma}^{*}(\br)\ \Phi_{q\sigma^{\prime}}(\br')}
{\omega+\omega_q-i 0_+}
\right\}
\een
where $q$ is a double index, representing a transition from
occupied KS orbital $i$ to unoccupied KS orbital $a$,
$\omega_{q\sigma}=\epsilon_{a\sigma}-\epsilon_{i\sigma}$, 
and $\Phi_{q\sigma}(\br) = \phi_{i\sigma}^*(\br)\phi_{a\sigma}(\br)$. $0_+$ means the limit as $0_+$ goes to zero from above (i.e., along the positive real axis). Thus $\chi\s$ is
completely determined by the ground-state KS potential.
It is the susceptibility of the non-interacting electrons sitting in the KS ground-state potential.\\

To relate the KS response to the true response, we examine how the KS potential in Eq. (\ref{vxc}) changes:
\ben
\delta v\sS(\br,t) = \delta v\extS(\br,t) + \delta v_{{\sss HXC}\sigma}(\bx) \ .
\een
Since $\delta v_{{\sss HXC}\sigma}(\bx)$ is due to an infinitesimal change in the density, it may be written in terms of its functional derivative, i.e., 
\ben
\delta v_{{\sss HXC}\sigma}(\bx) = \ssump\int d^3r'\int dt' \; f_{{\sss HXC}\sigma\sigma'}(\br\br',t-t') \; 
\delta \n\sigp(\br't') \ ,
\label{dvs}
\een
where 
\ben
\label{fxc}
f_{{\sss HXC}\sigma\sigma'}[\n_0](\br\br',t-t') =  \frac{\delta v_{{\sss HXC}\sigma}(\bx)}{\delta \n\sigp(\br' t')}
\bigg|_{\n_0} \ .
\een
The Hartree contribution is found by differentiating Eq. (\ref{hartpot}):
\bea
f\H[\n_0](\br\br',t-t') &=& \frac{\delta v\H(\bx)}{\delta\n\sigp(\br't')} \nonumber \\
						&=& \frac{\delta(t-t')}{|\br-\br'|}
\eea
while the remainder $f_{{\sss XC}\sigma\sigma'}[\n_0](\br\br',t-t')$ is known as the XC kernel.\\ 

By the definition of the KS potential, $\delta \n\sig(\bx)$ is
the same in both Eq. (\ref{chi}) and Eq. (\ref{chis}).  We can then insert Eq. (\ref{dvs}) into Eq. (\ref{chis}), equate
with Eq. (\ref{chi}) and solve for a general relation for any $\delta
\n\sig(\bx)$.  After Fourier transforming in time,
the central equation of TDDFT linear response\cite{PGG96} is a Dyson-like
equation for the true $\chi$ of the system:
\bea
\chi\dsig(\br\br'\omega)
&=& \chi_{{\sss S}\sigma\sigma'}(\br\br'\omega)
+ \sum_{\sigma_{1}\sigma_{2}}\int d^3r_1\int d^3r_2\ \chi_{{\sss S}\sigma\sigma_{1}}(\br\br_1\omega)\nonumber\\
&\times &
\left(\frac{1}{|\br_1-\br_2|}+f_{{\sss XC}\sigma_{1}\sigma_{2}} (\br_1\br_2\omega) \right)
\chi_{\sigma_{2}\sigma^{\prime}}(\br_2\br'\omega), \nonumber\\
\hspace{0.002cm}
\label{Dyson}
\eea

Notes:

\bei

\item

The XC kernel is a much simpler quantity than $v\xcS[\n](\bx)$, since
the kernel is a functional of only the ground-state density.

\item 
The kernel is
non-local in both space and time.  The non-locality in time
manifests itself as a frequency dependence in the Fourier transform,
$f_{{\sss XC}\sigma\sigma'}(\br\br'\omega)$.

\item
If $f\xc$ is set to zero in Eq. (\ref{Dyson}), physicists call it
the Random Phase Approximation (RPA).  The inclusion of $f\xc$ is an exactification
of RPA, in the same way the inclusion of $v\xc(\br)$ in ground-state
DFT was an exactification of Hartree theory.

\item
The Hartree kernel is instantaneous, i.e., local in time, i.e., has no memory, i.e., given exactly by an 
adiabatic approximation, i.e., is frequency independent.

\item
The frequency-dependent kernel is a very sophisticated object, since its frequency-dependence makes the solution of an RPA-type equation yield the exact $\chi$ (including all vertex corrections at every higher order term). It defies physical intuition and arguments based on the structure of the TDDFT equations are at best misleading. If any argument cannot be given in terms of many-body quantum mechanics, Eq. (\ref{Dyson}) cannot help.

\item
The kernel is, in general, complex, with real and imaginary parts related via Kramers-Kronig\cite{B06}.

\eei

Next, Casida\cite{C96} used ancient RPA technology, to produce equations in which the poles of
$\chi$ are found as the solution to an eigenvalue problem. The key is to expand in the basis
of KS transitions. We write $\delta\n\sig(\br t)$ as:

\ben
\delta\n\sig(\br\omega) = \sum_{q}\left( P_{q\sigma}(\omega)\Phi^{*}_{q\sigma}(\br) + 
P_{\bar{q}\sigma}(\omega)\Phi_{q\sigma}(\br)    \right)
\een
where $\bar{q} = (a,i)$ if $q = (i,a)$. This representation is used to solve Eq. (\ref{chis}) self-consistently using Eq. (\ref{dvs}), and yields two coupled matrix equations\cite{BA96}:
\ben
\left[\left(\begin{array}{cc} \mathbf{A} & \mathbf{B} \\ \mathbf{B}^{*} & \mathbf{A}^{*} \end{array}\right)-\omega\left(\begin{array}{cc} -\bbbone & 0 \\ 0 & \bbbone\end{array}\right)  \right]\left(\begin{array}{c}\mathbf{X} \\ \mathbf{Y}\end{array}\right) = - \left(\begin{array}{c}\mathbf{\delta v} \\ \mathbf{\delta v^{*}}\end{array}\right)
\een
where
$A_{q\sigma q'\sigma'} = \delta_{qq'}\delta\dsig\omega_{q\sigma} + K_{q\sigma q'\sigma'}$, $B_{q\sigma q'\sigma'} =  K_{q\sigma q'\sigma'}$, $X_{q\sigma} = P_{q\sigma}$, $Y_{q\sigma} = P_{\bar{q}\sigma}$  and
\ben
K_{q\sigma q'\sigma'}(\omega) = \int d\br\int d\br' \ \Phi_{q\sigma}(\br)~f_{{\sss HXC}\sigma\sigma'}(\br\br'\omega)~\Phi^{*}_{q'\sigma'}(\br') \ ,
\een
with 
\ben 
\delta v_{q\sigma}(\omega)  = \int d\br \ \Phi_{q\sigma}(\br)~\delta v\ext(\br\omega) \ .
\een
At an excitation energy, $\delta v = 0$  and choosing real KS orbitals and since
$(\mathbf{A}-\mathbf{B})$ is positive definite, we get:
\ben
\sum_{q'\sigma'} {\tilde\Omega}_{q\sigma q'\sigma'} (\omega)\ \vec{a}_{q'\sigma'} = \omega^2\ \vec{a}_{q\sigma},
\label{Casida}
\een
where
\ben
\tilde{\Omega} = (\mathbf{A}-\mathbf{B})^{1/2}(\mathbf{A}+\mathbf{B})(\mathbf{A}-\mathbf{B})^{1/2} \nonumber \ ,
\een
or 
\ben
{\tilde\Omega}_{q\sigma q'\sigma'}(\omega)  =  \omega^{2}_{q\sigma} \delta_{qq'}\delta_{\sigma\sigma'} + 2 \sqrt{\omega_{q\sigma}\omega_{q'\sigma'}}K_{q\sigma q'\sigma'} \ .
\label{Odef}
\een
Oscillator strengths $f_{q}$ may be calculated\cite{C96} from the normalized eigenvectors using 
\ben
f_{q\sigma} = \frac{2}{3}\left( |\vec{x}^{T}\mathbf{S}^{-1/2}\vec{a}_{q}|^{2}+|\vec{y}^{T}\mathbf{S}^{-1/2}\vec{a}_{q}|^{2}+|\vec{z}^{T}\mathbf{S}^{-1/2}\vec{a}_{q}|^{2}  \right) \ ,
\een
where
\ben
S_{qq'} = \delta_{qq'}\delta\dsig / w_{q'} \nonumber \ .
\een

\begin{figure}[htb]
\unitlength1cm
\begin{picture}(12,6.9)
\put(-6.5,-4){\makebox(12,6){
\includegraphics{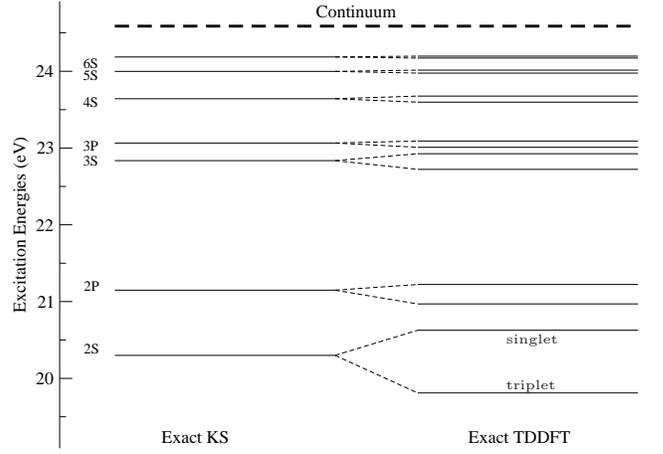}
}}
\put(6.7,1){\tiny{triplet}}
\put(6.7,1.6){\tiny{singlet}}
\end{picture}
\caption{Transitions for the Helium atom using
in ground-state DFT on the left, and TDDFT on the right.  In both cases,
the exact functionals have been used. The results for employing the exact XC kernel in TDDFT linear response are known from calculations using Ref. \cite{S26}. In each pair of lines on the right, the triplet is the lower.}
\label{f:he_ex}
\end{figure}

Figure \ref{f:he_ex} shows the results of {\em exact} DFT calculations
for the He atom.  On the left, we consider just transitions between 
the exact ground-state KS occupied (1s) to unoccupied orbitals.
These are {\em not} the true excitations of the system, nor are they
supposed to be.  However, applying TDDFT linear response theory, using
the exact kernel on the exact orbitals, yields the exact excitations of
the He atom.  Spin-decomposing produces both singlet and triplet excitations.

\subsection{Approximations}
\label{s:tdappr}
As in the ground-state case, while all the equations above are formally exact, a practical TDDFT
calculation requires an approximation for the unknown XC potential.\\
The most common approximation in TDDFT is the {\em adiabatic}
approximation, in which
\ben
v_{{\sss XC}\sigma}\adia [\n](\br t) = v_{{\sss XC}\sigma}^{\rm gs} [\n_0] (\br)|_{\n_{0\sigma}(\br)=\n\sig(\br t)},
\label{adia}
\een
i.e., the XC potential at any time is simply the ground-state
XC potential at that instant.
This obviously becomes
exact for sufficiently slow perturbations in time, 
in which the system always stays in its
instantaneous ground-state. Most applications, however, are not in this 
slowly varying regime. Nevertheless, results obtained within the adiabatic
approximation are remarkably accurate in many cases. 

Any ground-state
approximation (LDA, GGA, hybrid) automatically provides an adiabatic
approximation for use in TDDFT.
The most famous is the adiabatic local density
approximation (ALDA). It employs the functional form of the static LDA with a
time-dependent density:
\ben
\label{ALDA}
v_{{\sss XC}\sigma}\ALDA[n](\br t) =
v\xc\unif(\n_{\alpha}(\br t),\n_{\beta}(\br t)) =
\frac{d e\xc\unif}{d \n\sig}
\bigg|_{\n\sig = \n\sig(\br t)} \,.
\een
Here $e\xc\unif(\n_{\alpha},\n_{\beta})$ is the accurately
known exchange-correlation energy density of the
uniform electron gas of spin densities $\n_{\uparrow},\n_{\downarrow}$.
For the time-dependent exchange-correlation
kernel of  Eq. (\ref{fxc}),  Eq. (\ref{ALDA}) leads to
\ben
\label{fxcALDA}
f_{{\sss XC}\sigma\sigma'}\ALDA[n_0](\bx,\bx') =
\delta^{(3)}(\br-\br')\, \delta(t-t')\,
\frac{d^2 e\xc\unif}{d\n\sig d\n\sigp}
\bigg|_{\n\sig = \n_{0\sigma}({\bf r})} \,.
\een
The time Fourier-transform of the kernel
has no frequency-dependence at all in any adiabatic approximation.
Via a Kramers-Kronig relation, this implies that
it is purely real\cite{B06}.

Thus, any TDDFT linear response calculation can be considered as occuring in two
steps:

\begin{enumerate}

\item
An approximate ground-state DFT calculation is done, and a self-consistent
KS potential found. Transitions from occupied to unoccupied KS orbitals provide zero-order approximations to the optical excitations.

\item
An approximate TDDFT linear response calculation is done on the orbitals
of the ground-state calculation. This corrects the KS transitions into the true optical transitions.

\end{enumerate}

In practice both these steps have errors built into them. 

\section{Implementation and basis sets}
\label{s:imp}
In this section we discuss how TDDFT is implemented numerically. TDDFT
has the ability to calculate many different quantities and different
techniques are sometimes favored for each type. For some purposes, e.g., if strong fields are present, 
it can be better to propagate forward in time the KS orbitals
using a real space grid\cite{VOC99,MCBR03} 
or with plane 
waves\cite{CPMD95}.
For finite-order response,
Fourier transforming to frequency space with localized basis functions
may be preferable\cite{Furche01a}.  Below, we discuss in detail how this approach works,
emphasizing the importance of basis-set convergence.

\subsection{Density matrix approach}
\label{s:dma}
Instead of using orbitals, we can write the dynamics of the TDKS
systems in terms of the one-particle density matrix
$\gamma_{\sigma}(\br \br' t)$ of
the TDKS determinant. $\gamma_{\sigma}(\br \br' t)$ has the spectral representation
\begin{equation}
  \label{eq:spec}
  \gamma_{\sigma}(\br \br' t) = \sum_{j=1}^N \phi_{j\sigma}(\br t)
  \phi_{j\sigma}^*(\br' t),
\end{equation}
i.e., the $N_{\sigma}$ TDKS orbitals are the eigenfunctions of $\gamma_{\sigma}$. The
eigenvalue of all TDKS orbitals, which is their \emph{occupation
  number}, is always $1$, which reflects the fact that the TDKS system
is non-interacting. Equivalently, $\gamma_{\sigma}$ satisfies the idempotency
constraint
\begin{equation}
  \label{eq:idemp}
  \gamma_{\sigma}(\br \br' t) = \int dx_1 \gamma_{\sigma}(\br \br_{1}
  t)\gamma_{\sigma}(\br_{1} \br' t).
\end{equation}
The normalization of the TDKS orbitals implies that the trace of
$\gamma_{\sigma}$ be $N_{\sigma}$.

Using the TDKS equations \eqref{TDKS}, one finds that the
time-evolution of $\gamma_{\sigma}$ is governed by the von-Neumann equation
\begin{equation}
  \label{eq:eom}
  i\frac{\partial}{\partial t} \gamma_{\sigma}(t) = \bigl[ H_{\sigma}[n](t),
  \gamma_{\sigma}(t) \bigr],
\end{equation}
where $H_{\sigma}[n](\bx) = -\nabla^2/2 + v_{s\sigma}[n](\bx)$ is the
TDKS one-particle 
Hamiltonian. Although $\gamma_{\sigma}$ has no direct physical meaning, it
provides the interacting density and current density: The density is simply
\begin{equation}
  n_{\sigma}(\br t) = \gamma_{\sigma}(\br \br t),
\end{equation}
and the current density can be obtained from
\begin{equation}
  \mathbf{j}_{\sigma}(\br t) = \left. \frac{1}{2i}\bigl(
      \nabla_{\mathbf{r}}-\nabla_{\mathbf{r'}}\bigr)
      \gamma_{\sigma}(\br \br' t) 
    \right|_{\br'=\br}.
\end{equation}
Thus, one can either propagate the TDKS orbitals using the TDKS
equations \eqref{TDKS}, or equivalently one can propagate the TDKS
one-particle density matrix $\gamma_{\sigma}$ using the von-Neumann equation
\eqref{eq:eom}, subject to the idempotency contstraint
\eqref{eq:idemp} and normalized to $N_{\sigma}$. 

In practice, it is often preferable to use $\gamma_{\sigma}$ instead of the
TDKS orbitals. $\gamma_{\sigma}$ is unique (up to a gauge transformation),
while the orbitals can be mixed arbitrarily by unitary
transformations. Both $n_{\sigma}$ and $\mathbf{j}_{\sigma}$ are linear in $\gamma_{\sigma}$,
while they are quadratic in the orbitals; also, the TDKS equations are
inhomogeneous in the orbitals due to the density dependence of $H_{\sigma}$,
while they are homogeneous in $\gamma_{\sigma}$. A response theory based on the
TDKS density matrix is therefore considerably simpler than one based
on the orbitals. Finally, the use of $\gamma_{\sigma}$ is computationally more
efficient than using orbitals\cite{Furche01a}.

\subsection{Basis Sets}

In response theory, the basis functions $\chi_{\nu}(\br)$ are usually
chosen to be time-independent; for strong fields or coupled
electron-nuclear dynamics, time-dependent basis functions can be more
appropriate.

\subsection{Naphthalene converged}

Table \ref{tab:bascv_ener} shows the basis set convergence of the
first six singlet excitation energies of naphthalene computed using
the PBE XC functional; the corresponding oscillator
strengths for some of the transitions are also given. Similar basis-set convergence studies on small
model systems should precede applications to large systems. In
practice, the systems and states of interest, the target accuracy, the
methods used, and the computational resources available will determine which
basis set is appropriate.

\begin{table*}
\caption{\label{tab:bascv_ener} Basis set convergence of the first six
singlet excitation energies (in eV) and oscillator strengths (length
gauge) of naphthalene. The basis set
acronyms are defined in the text. $N_{\text{bf}}$ denotes the number
of Cartesian basis functions, and CPU denotes the CPU time (seconds) on a single
processor of a 2.4 GHz Opteron Linux workstation. The PBE functional was used
for both the ground-state and TDDFT calculations. The ground-state
structure was optimized at the PBE/TZVP/RI-level. Experimental results
were taken from Ref. \cite{Rubio94a}.}
\begin{ruledtabular}
\begin{tabular}{c|cccccccc}
Basis set & 1$\ ^1B_{3u}$ & 1$\ ^1B_{2u}$  (Osc. Str.) & 2$\ ^1A_g$ & 1$\ ^1B_{1g}$
& 2$\ ^1B_{3u}$ (Osc. Str.) & 1$\ ^1A_u$ & $N_{\text{bf}}$ & CPU \\
  \hline
  SV 		& 4.352 & 4.246  (0.0517)& 6.084 & 5.254 & 5.985  (1.1933) & 6.566 & 106 & 24 \\
  SV(P) 		& 4.272 & 4.132  (0.0461)& 5.974 & 5.149 & 5.869  (1.1415) & 6.494 & 166 & 40 \\
  6-31G$^*$ 	& 4.293 & 4.154   \ \ \ \ \ \ \ \ \ \ \ \  & 6.021 & 5.185 & 5.902 \ \ \ \ \ \ \ \ \ \ \ \ & 7.013 & 166 & 40 \\
  SVP 		& 4.262 & 4.125  (0.0466)& 5.960 & 5.136 & 5.852   (1.1402) & 6.505 & 190 & 48 \\
  aug-SV(P)	 & 4.213 & 4.056 (0.0417)& 5.793 & 4.993 & 5.666  (1.1628) & 5.338 & 266 & 168 \\
  TZVP 		& 4.209 & 4.051  (0.0424) & 5.834 & 5.030 & 5.715  (1.1455) & 6.215 & 408 & 408 \\
  TZVPP 		& 4.208 & 4.050  (0.0425) & 5.830 & 5.027 & 5.711  (1.1464) & 6.231 & 480 & 568 \\
  cc-pVTZ 	& 4.222 & 4.064  (0.0427) & 5.870 & 5.061 & 5.747  (1.1355) & 6.062 & 470 & 528 \\
  aug-TZVP 	& 4.193 & 4.031  (0.0407) & 5.753 & 4.957 & 5.622  (1.1402) & 5.141 & 608 & 2000 \\
  aug-TZVP/RI & 4.193 & 4.031  (0.0407)& 5.752 & 4.957 & 5.621  (1.1401) & 5.142 & 608 & 400 \\
  QZVP 		& 4.197 & 4.036  (0.0416) & 5.788 & 4.989 & 5.667  (1.1569) & 5.672 & 1000 & 6104 \\
  aug-QZVP 	& 4.192 & 4.029  (0.0406) & 5.748 & 4.954 & 5.616  (1.1330) & 5.071 & 1350 & 28216 \\
  \hline
   expt. & 3.97, 4.0 & 4.45, 4.7 (0.102, 0.109) & 5.50, 5.52 & 5.28, 5.22 & 5.63, 5.55, 5.89  (1.2, 1.3) & 5.6 \\
\end{tabular}
\end{ruledtabular}
\end{table*}

With a model small molecule, we can find the basis-set convergence limit of a method. Both excitation energies and oscillator strengths are essentially
converged within the aug-QZVP basis set. QZVP stands for a quadruple
zeta valence basis set with polarization functions\cite{Weigend03a}, and the prefix
aug- denotes additional diffuse functions on non-hydrogen
atoms, which were taken from Dunning's aug-cc-pVQZ basis set\cite{Kendall92a}. For C
and H, this corresponds to $[8s5p4d3f2g]$ and $[4s3p2d1f]$,
respectively, where the numbers in brackets denote shells of contracted
Gaussian type orbitals (CGTOs), as usual. We will take the aug-QZVP results
as a reference to assess the effect of smaller basis sets.

\subsection{Double zeta basis sets}

The smallest basis in Table \ref{tab:bascv_ener} is of split valence (SV)
or double zeta valence quality\cite{Schaefer92a}, without polarization functions. This
basis set consists of two CGTOs per valence orbital and one per core
orbital, i.e. $[3s2p]$ for C and $[2s]$ for H. Another popular double zeta
valence basis set is 6-31G \cite{Hehre72a}. The SV basis set can be used to obtain a very
rough qualitative description of the lowest valence excited states only,
e.g. 1$\ ^1B_{3u}$ and 1$\ ^1B_{2u}$. Higher and diffuse excitations,
such as 1$\ ^1A_u$, are much too high in energy or can be missed completely
in the spectrum. Since unpolarized basis sets also give poor results
for other properties such as bond lengths and energies, their use is
generally discuraged nowadays.

\subsection{Polarization functions}

By adding a single set of polarization functions to non-hydrogen
atoms, the SV results for valence excitations can be considerably
improved, at still moderate computational cost. The
resulting basis set is termed SV(P) and consists of $[3s2p1d]$ for C
and $[2s]$ for H\cite{Schaefer92a}. The basis set errors in the first two valence
excitation energies is reduced by about 50\%. There is also a dramatic
improvement in the oscillator strength of the dipole allowed
transitions. This is expected from the
limiting case of a single atom, where the first dipole allowed
transition from a valence shell of $l$ quantum number $l_v$ generally
involves orbitals with $l$-quantum number $l_v+1$. Basis sets of SV(P)
or similar quality are often the first choice for TDDFT applications
to large systems, especially if only the lowest states are of interest
and/or diffuse excitations are quenched, e.g. due to a polar
environment. The popular 6-31G$^*$ basis set \cite{Hehre72a,Hariharan73a} has essentially the same size as SV(P)
but performs slightly poorer in our example. 

Adding a single set of $p$ type polarization functions to hydrogen atoms
produces the SVP basis set\cite{Schaefer92a}. These functions mainly describe C-H $\sigma^*$
type excitation in molecules which usually occur in the far UV and are
rarely studied in applications. In our example, going from SV(P) to
SVP has no significant effect. This may be different for molecules
containing strongly polarized hydrogen-element or hydrogen bridge
bonds. 

Next, aug-SV(P) is an SV(P) basis set augmented by a $[1s1p1d]$ set of
primitive Gaussians with small exponents (from Dunning's aug-cc-pVDZ
\cite{Kendall92a}), often called ``diffuse
functions''. As shown in Table \ref{tab:bascv_ener}, the effect of
diffuse augmentation is a moderate downshift of less than 0.1 eV for
the first two singlet excitation energies. This behavior is typical of
lower valence excited states having a similar extent as the ground-state. Our example also shows that diffuse functions can have a
significant effect on higher excitations. An extreme case is the 1$\
^1A_u$ state which is an excitation into the $10au$ orbital having the
character of a $3s$ Rydberg state (of the entire molecule). The
excitation energy of this state is lowered by more than 1 eV upon
diffuse augmentation.

While polarization functions are necessary for a qualitatively correct
description of transition dipole moments, additional diffuse
polarization functions can account for radial nodes in the first-order
KS orbitals, which further improves computed transition moments and
oscillator strengths. These benefits have to be contrasted with a
significant increase of the computational cost: In our example, using
the aug-SV(P) basis increases the computation time by about a factor of
4. In molecules with more than 30-40 atoms, most excitations of
interest are valence excitations, and the use of diffuse augmentation
may become prohibitively expensive because the large extent of these
functions confounds integral prescreening.

\subsection{Triple zeta basis sets}

In such cases, triple zeta valence (TZV) basis sets can be a better
alternative. The TZVP (def-2-TZVP, Ref. \cite{Weigend05a}) basis set
corresponds to $[5s3p2d1f]$ on C and $[3s1p]$ on H. It provides
a description of the valence electrons that is quite accurate for many
purposes when density functionals are used. At the same time, there is
a second set of polarization functions on non-hydrogen atoms. The
excitation energies of valence states are essentially converged in
this basis set, see Table \ref{tab:bascv_ener}. However, diffuse
states are too high in energy. There is very little change going to
the TZVPP basis, which differs from TZVP only by an additional set of
polarization functions on H. Dunning's cc-pVTZ basis set
\cite{Dunning89a} performs similar to TZVP and TZVPP. However, Dunning
basis sets are based on a generalized contraction scheme for valence
orbitals, as opposed to the segmented contracted SV, TZV and QZV basis
sets. The latter are more efficient for larger systems, because more
integrals vanish. 

\subsection{Diffuse functions}

Adding a $[1s1p1d1f]$ set of diffuse functions to TZVP we obtain the
aug-TZVP basis set. The aug-TZVP excitation energies of all states
except the 1$\ ^1A_u$ Rydberg state are within 0.01 eV of the
reference aug-QZVP results and can be considered essentially converged
for the purposes of present TDDFT. A similar observation can be made
for the oscillator strengths in Table \ref{tab:bascv_ener}. 

Going to
the even larger quadruple zeta valence (QZV) basis sets, the results
change only marginally, but the computation times increase
substantially. In density functional theory, these basis sets are
mainly used for bechmarks and calibration.

\subsection{Resolution of the identity}
For comparison, we have included results that were obtained using the
resolution of the identity approximation for the Coulomb energy
(RI-J) \cite{Mintmire82a,Eichkorn95a}. It is obvious that the error introduced by the RI-J
approximation is much smaller than the basis set error, while the
computation time is reduced by a factor of 5. The RI-J approximation
is so effective because the computation of the Coulomb (Hartree)
energy and its respose is the bottleneck in conventional (TD)DFT
calculations. RI-J replaces the four-index Coulomb integrals by
three-index and two-index integrals, which considerably lowers the
algorithmic pre-factor\cite{RF05}. It is generally safe to use with the
appropriate auxiliary basis sets. As soon as hybrid functionals are
used, however, the computation of the exact exchange becomes
rate-determining.

\subsection{Summary}
To summarize, for larger molecules, SV(P) or similar basis sets are
often appropriate due to their good cost-to-performance ratio. We
recommend to check SV(P) results by a TZVP calculation whenever
possible. Diffuse functions should be used sparingly for molecules
with more than about 20 atoms.

\section{Performance}
\label{s:perf}

This chapter is devoted to studying and analyzing the performance of TDDFT, assuming
basis-set convergence.  We dissect many of the sources of error in 
typical TDDFT calculations.

To get an overall impression, 
  a small survey is given
by Furche and Ahlrichs\cite{FA02}.  Typical chemical
calculations are done with the B3LYP\cite{B93} functional, and
typical results are transition
frequencies within 0.4 eV of experiment, and structural properties
of excited states 
are almost as good as those of ground-state calculations
(bond lengths to within 1\%,
dipole moments to within 5\%, vibrational frequencies to within 5\%).
Most importantly, this level of accuracy appears sufficient in
most cases to qualitatively identify the nature of the most intense
transitions, often debunking cruder models that have been used for
interpretation for decades.  This is proving especially useful for
the photochemistry of biologically relevant molecules\cite{MLVC03}

\subsection{Example: Napthalene Results}

As an illustration, compare the performance of various density functionals and
wavefunction methods for the first singlet excited states of
naphthalene in Tables \ref{tab:dft_method_ener}, \ref{tab:wf_method_ener} and
\ref{tab:method_osc}. All calculations were performed using the
aug-TZVP basis set; the complete active space SCF with second-order
perturbation theory (CASPT2) results from Ref. \cite{Rubio94a} were
obtained in a smaller double zeta valence basis set with some diffuse
augmentation. The experimental results correspond to band maxima from
gas-phase experiments; however, the position of the band maximum does
not necessarily coincide with the vertical excitation energy,
especially if the excited state structure differs significantly from
the ground-state structure. For the lower valence states, the CASPT2
results can therefore be expected to be at least as accurate as the
experimental numbers. For higher excited states, the basis set used in
the CASPT2 calculations appears rather small, and the approximate
second-order coupled cluster values denoted RICC2 \cite{Christiansen95a,Haettig00a,Haettig02a} might be a better
reference.  Thus our best guess (denoted ``best" in the Tables) is
from experiment for the first 4 transitions, CASPT2 for the 5th, and
RICC2 for the 6th.

\begin{table}
\caption{\label{tab:dft_method_ener} Performance of various density
  functionals for the first six
  singlet excitation energies (in eV) of naphthalene. An aug-TZVP basis
  set and the PBE/TZVP/RI ground-state structure was used.
The ``best" estimates of the true excitations were from experiment
and calculations, as described in text.}
\begin{ruledtabular}
\begin{tabular}{c|cccccccc}
Method & 1$\ ^1B_{3u}$ & 1$\ ^1B_{2u}$ & 2$\ ^1A_g$ & 1$\ ^1B_{1g}$
& 2$\ ^1B_{3u}$ & 1$\ ^1A_u$ \\ 
\multicolumn{7}{c}{Pure density functionals}\\
     LSDA  & 4.191 & 4.026 & 5.751 & 4.940 & 5.623 & 5.332 \\
     BP86  & 4.193 & 4.027 & 5.770 & 4.974 & 5.627 & 5.337 \\
     PBE   & 4.193 & 4.031 & 5.753 & 4.957 & 5.622 & 5.141 \\
\hline
\multicolumn{7}{c}{Hybrids}\\
     B3LYP & 4.393 & 4.282 & 6.062 & 5.422 & 5.794 & 5.311 \\
     PBE0  & 4.474 & 4.379 & 6.205 & 5.611 & 5.889 & 5.603 \\
\hline
     ``best". &  4.0 & 4.5 & 5.5 & 5.5 &5.5 & 5.7\\
\end{tabular}
\end{ruledtabular}
\end{table}

\begingroup
\squeezetable
\begin{table}

\caption{\label{tab:wf_method_ener} Performance of various 
  wavefunction methods for the excitations of Table I.  
  The  aug-TZVP basis
  set and the PBE/TZVP/RI ground-state structure was used for all
  except the CASPT2 results, which were taken from
  Ref. \cite{Rubio94a}. Experimental results are also from
  Ref. \cite{Rubio94a}.}
\begin{ruledtabular}
\begin{tabular}{c|cccccccc}
Method & 1$\ ^1B_{3u}$ & 1$\ ^1B_{2u}$ & 2$\ ^1A_g$ & 1$\ ^1B_{1g}$
& 2$\ ^1B_{3u}$ & 1$\ ^1A_u$ \\ 
     CIS   & 5.139 & 4.984 & 7.038 & 6.251 & 6.770 & 5.862 \\
     CC2   & 4.376 & 4.758 & 6.068 & 5.838 & 6.018 & 5.736 \\
    CASPT2 & 4.03  & 4.56  & 5.39  & 5.53  & 5.54  & 5.54  \\
\hline
\multirow{2}{*}{expt.} & \multirow{2}{*}{3.97, 4.0} & \multirow{2}{*}{4.45, 4.7} & \multirow{2}{*}{5.50, 5.52} & \multirow{2}{*}{5.28, 5.22} & 5.63,5.55 &\\
								&	&			&			&			&	5.89 & \\
\hline
     ``best". &  4.0 & 4.5 & 5.5 & 5.5 &5.5 & 5.7\\
\end{tabular}
\end{ruledtabular}
\end{table}
\endgroup

\begin{table}
\caption{\label{tab:method_osc} Performance of various density
  functionals and correlated wavefunction methods for the oscillator
  strengths of the first three dipole-allowed transitions of
  naphthalene. A aug-TZVP basis   set and the PBE/TZVP/RI ground-state
  structure was used for all except the CASPT2 results, which were
  taken from Ref. \cite{Rubio94a}.} 
\begin{ruledtabular}
\begin{tabular}{c|ccc}
  Method & 1$\ ^1B_{3u}$ & 1$\ ^1B_{2u}$ & 2$\ ^1B_{3u}$  \\
  \hline
  LSDA  & 0.0000 & 0.0405 & 1.1517 \\
  BP86  & 0.0000 & 0.0411 & 1.1552 \\
  PBE   & 0.0000 & 0.0407 & 1.1402 \\
  B3LYP & 0.0000 & 0.0539 & 1.2413 \\
  PBE0  & 0.0000 & 0.0574 & 1.2719 \\
LHF/LSDA& 0.0000 & 0.0406 & 1.2089 \\
LHF/PBE & 0.0000 & 0.0403 & 1.2008 \\
  CIS   & 0.0002 & 0.0743 & 1.8908 \\
  CC2   & 0.0000 & 0.0773 & 1.4262 \\
  CASPT2& 0.0004 & 0.0496 & 1.3365 \\
  expt. & 0.002  & 0.102, 0.109 & 1.2, 1.3 \\

\end{tabular}
\end{ruledtabular}
\end{table}

We begin with some general observations.

\bei

\item The excitation energies
predicted by the GGA functionals BP86 and PBE differ only marginally
from the LSDA results (an exception being the 1$\ ^1A_u$ Rydberg
state, whose PBE excitation energy is substantially lower than those
of all other methods). Note however that GGA functionals generally
improve over LSDA results for other excited state properties such as
structures or vibrational frequencies.

\item  Hybrid mixing leads to
systematically higher excitation energies. On average, this is an
improvement over the GGA results which are systematically too
low. However, while diffuse excitations benefit from hybrid mixing due
to a reduction of self-interaction error, valence excitation energies
are not always improved, as is obvious for the 1$\ ^1B_{3u}$ and 2$\
^1B_{3u}$ valence states.

\item
The 1$\ ^1B_{2u}$ state is erroneously
predicted below the 1$\ ^1B_{3u}$ state by all density functionals,
which is a potentially serious problem for applications in
photochemistry; this is not corrected by hybrid mixing.

\item The configuration-interaction singles (CIS) method 
which uses a Hartree-Fock reference that is computationally as expensive as hybrid TDDFT produces errors
that are substantially larger, especially for valence states.
The coupled cluster and CASPT2 methods are far more expensive, and
scale prohibitively as the system size grows.

\eei

The 1$\ ^1B_{2u}$ excitation is polarized along the short axis of the
naphthalene molecule. In Platt's nomenclature of excited states of
polycyclic aromatic hydrocarbons (PAHs), 1$\ ^1B_{2u}$ corresponds to the
$^1L_a$ state. This state is of more ionic character than the 1$\
^1B_{3u}$ or $^1L_b$ state. Parac and Grimme have
pointed out \cite{Parac03a} that GGA functionals considerably
underestimate the excitation energy of the $^1L_a$ state in PAHs. This
agrees with the observation that the 1$\ ^1B_{2u}$ excitation of
naphthalene is computed 0.4-0.5 eV too low in energy by LSDA and GGA
functionals, leading to an incorrect ordering of the first two singlet
excited states.

\subsection{Influence of the ground-state potential}
\label{s:gspot}

From the very earliest calculations of transition frequencies\cite{PGG96,C96},
it was recognized that the inaccuracy of
standard density functional approximations (LDA, GGA, hybrids)
for the ground-state XC potential
leads to inaccurate KS eigenvalues. 
Because the approximate KS potentials have incorrect asymptotic behavior
(they
decay exponentially, instead of as $-1/r$, as seen in Fig. \ref{f:LDA_vs_exKS}), the KS orbital
eigenvalues are insufficiently negative, the ionization
threshold is far too low, and Rydberg states are often
unbound.  

Given this disastrous behavior, many 
methods have been developed to asymptotically correct potentials\cite{GGGB02,WAY03}.
Any corrections to the ground-state potential are
dissatisfying, however, as the resulting potential is {\em not} a functional
derivative of an energy functional.  Even mixing one approximation
for $v\xc(\br)$ and another for $f\xc$ has become popular.
A more satisfying route is to use the optimized effective potential (OEP) 
method\cite{GKKG98,UGG95}
and include exact exchange or other self-interaction-free functionals\cite{DG03}.
This produces a far more accurate
KS potential, with the correct asymptotic behavior.
The chief remaining error is simply the correlation contribution to the
position of the HOMO, i.e, a small shift.  All the main features
below and just above $I$ are retained.\\

\subsubsection{N$_{2}$, a very small molecule}

\begin{table}
\caption{\label{tab:ks_nrg_n2}Orbital energies of the KS energy levels for N$_{2}$ at separation $R=2.0744$a.u. Orbitals calculated
 using the LDA potential are shown for two different numerical methods. The first is fully numerical basis set free while the
  other uses the Sadlej ($52$ orbitals) basis set\cite{Sb91}. The OEP method uses the 
  EXX (KLI) approximation and is also calculated basis set free.}
\begin{ruledtabular}
\begin{tabular}{lccc}
\multicolumn{4}{c}{Energies in (eV)}\\
Orbital & LDA& LDA  & OEP\footnotemark[1]\\
        & basis set free\footnotemark[1] 	& Sadlej\footnotemark[2]  & \\
\hline
\multicolumn{4}{l}{Occupied orbitals}  \\
$1\sigma_{g}$ & -380.05	& -380.82	& -391.11   \\
$1\sigma_{u}$ & -380.02	& -380.78	& -391.07   \\
$2\sigma_{g}$ & -28.24		& -28.52		& -35.54    \\
$2\sigma_{u}$ & -13.44		& -13.40		& -20.29    \\
$1\pi_{u}$    	& -11.89		& -11.86		& -18.53    \\
$3\sigma_{g}$ & -10.41		& -10.38		& -17.15    \\
\multicolumn{4}{l}{Unoccupied orbitals}\\
$1\pi_{g}$    & -2.21	& -2.23		& -8.44     \\
$4\sigma_{g}$ & -0.04	& 0.66		& -5.05     \\
$2\pi_{u}$    & $>0$	& 1.93		& -4.04     \\
$3\sigma_{u}$ & $>0$	& 1.35		& -3.54     \\
$1\delta_{g}$ & $>0$	& - 			& -2.76     \\
$5\sigma_{g}$ & $>0$	& 3.20		& -2.49     \\
$6\sigma_{g}$ & $>0$	& -			& -2.33     \\
$2\pi_{g}$    & $>0$	& 3.89		& -2.17     \\
$3\pi_{u}$    & $>0$	& -			& -2.04     \\

\end{tabular}
\end{ruledtabular}
\footnotemark[1]{From Ref.\cite{GPG00}.}
\footnotemark[2]{from Ref.\cite{JCS96}.}
\end{table}

\begingroup
\squeezetable
\begin{table}
\caption{\label{tab:n2_trans}Comparison of the vertical excitation energies for the first twelve excited states of N$_{2}$ calculated using
 different methods for the SCF step. In all cases the KS orbitals from the SCF step are inputed into Casida's equations with the ALDA XC
  kernel. For the LDA calculated with the Sadlej basis set, the bare KS transition frequencies are given to demonstrate how they are 
  corrected towards their true values using Casida's equations. Also given are the mean absolute errors for each method, errors in backets
   are calculated for the lowest eight transitions only.}
\begin{ruledtabular}
\begin{tabular}{cccccccc}
\multicolumn{8}{c}{Excitation energy (eV)}
\\
State & Excitation & BARE KS\footnotemark[1] & ALDA\footnotemark[1] & ALDA\footnotemark[2] & LB$94$\footnotemark[3] & OEP\footnotemark[4] & Expt\footnotemark[5] \\
\hline
\multicolumn{8}{c}{Singlet $\rightarrow$ singlet transitions}\\
w$^{1}\Delta_{u}$		& $1\pi_{u}\rightarrow 1\pi_{g}$	&9.63	&10.20	&10.27  &9.82	 &10.66	 &10.27 \\
a$^{\prime1}\Sigma^{-}_{u}$	& $1\pi_{u}\rightarrow 1\pi_{g}$  	&9.63	&9.63	&9.68   &9.18	 &10.09	 &9.92 \\
a$^{1}\Pi_{g}$			& $3\sigma_{g}\rightarrow 1\pi_{g}$	&8.16	&9.04	&9.23   &8.68	 &9.76	 &9.31 \\
a$^{\prime\prime1}\Sigma_{g}$ 	& $3\sigma_{g}\rightarrow 4\sigma_{g}$  &-	&-	&10.48  &-	 &12.47	 &12.20 \\ 
o$^{1}\Pi_{u}$ 			& $2\sigma_{u}\rightarrow 1\pi_{g}$	&-	&-	&13.87	&-	 &14.32	 &13.63 \\
c$^{1}\Pi_{u}$ 			& $3\sigma_{g}\rightarrow 2\pi_{u}$	&-	&-	&11.85	&-	 &13.07	 &12.90 \\
\multicolumn{8}{c}{Singlet $\rightarrow$ triplet transitions}\\
C$^{3}\Pi_{u}$			& $2\sigma_{u}\rightarrow 1\pi_{g}$	&11.21	&10.36	&10.44  &10.06	 &11.05	 &11.19 \\
B$^{\prime3}\Sigma^{-}_{u}$	& $1\pi_{u}\rightarrow 1\pi_{g}$        &9.63	&9.63	&9.68   &9.18	 &10.09	 &9.67 \\
W$^{3}\Delta_{u}$		& $1\pi_{u}\rightarrow 1\pi_{g}$        &9.63	&8.80	&8.91   &8.32	 &9.34	 &8.88 \\
B$^{3}\Pi_{g}$			& $3\sigma_{g}\rightarrow 1\pi_{g}$	&8.16	&7.50	&7.62   &7.14	 &8.12	 &8.04 \\
A$^{3}\Sigma^{+}_{u}$		& $1\pi_{u}\rightarrow 1\pi_{g}$        &9.63	&7.84	&8.07   &7.29	 &8.51	 &7.74 \\
E$^{3}\Sigma_{g}^{+}$   	& $3\sigma_{g}\rightarrow 4\sigma_{g}$  &-	&-	&10.33	&12.32	 &11.96	 &12.00 \\
\hline
\multicolumn{2}{c}{Mean Absolute Error} &(0.61)  &(0.27)   &0.54  &(0.63)  &0.34  &  
\end{tabular}
\end{ruledtabular}
\footnotetext[1]{Using Sadlej basis set. From Ref.\cite{JCS96}.}
\footnotetext[2]{Basis set free. From Ref.\cite{GPG00}.}
\footnotetext[3]{From Ref.\cite{CJCS98}.}
\footnotetext[4]{Using KLI approximation. From Ref.\cite{GPG00}.}
\footnotetext[5]{Computed in \cite{OGD85} from the spectroscopic constants 
of Huber and Herzberg \cite{HH79}.}
\end{table}
\endgroup

A simple system to see the effect of the various ground-state potentials is the N$_{2}$ molecule. In all the cases discussed below, 
a SCF step was carried out using the ground-state potential to find the
KS levels. These are then used as input to Eq. (\ref{Casida}) with the ALDA XC kernel.\\

In Table \ref{tab:ks_nrg_n2}, the KS energy levels for the LDA functional are shown. It is very clear to see that the eigenvalues 
for the higher unoccupied states are positive.  As mentioned this is due to
the LDA potential being too shallow and not having the correct asymptotic behavior. Comparing the basis-set calculation with the 
basis-set-free calculation, the occupied orbitals are in good agreement. 
However for the unoccupied states that are unbounded in LDA, basis sets cannot describe these correctly and give a positive energy value which can vary greatly from one basis set to another.\\

In Table \ref{tab:n2_trans}, the bare KS transition frequencies between these levels are shown. Note that they are in rough 
agreement with the experimental values and that they lie inbetween the
singlet-singlet and singlet-triplet transitions\cite{SUG98}. The ALDA XC kernel $f\xc\ALDA$ then shifts the KS transitions towards their 
correct values. For the eight lowest transitions LDA does remarkably
well, the mean absolute error (MAE) being $0.27\rm$eV for the Sadlej basis set. For higher transitions it fails drastically, the 
MAE increases to $0.54\rm$eV when the next four transitions are
included. This increase  in the MAE is attributed to a cancellation of errors that lead to good frequencies for the lower 
transitions\cite{GPG00}. Since LDA binds only two unoccupied orbitals, 
it cannot accurately describe transitions to higher orbitals. In basis set calculations, the energies of the unbound orbitals 
which have \emph{converged} will vary wildly and cannot give trusted
transition frequencies.\\

One class of XC functionals that would not have this problem are the asymptotically corrected (AC) functionals
\cite{GGGB02,WAY03,TH98,WY03,CS00}. 
LB$94$\cite{LB94} is one
 such of these and its performance is shown in
Table \ref{tab:n2_trans}. AC XC potentials tend to be too shallow in the core region, so the KS energy levels will be too low while
 the AC piece will force the higher KS states to be bound 
and their energies will cluster below zero. Thus it is expected that using AC functionals will consistently underestimate the transitions frequencies.\\

A much better approach is using the OEP method. The KS orbitals found using this method are self-interaction free and are usually better approximations to the true KS orbitals. 
OEP will also have the correct asymptotic behavior and as we can see in Table \ref{tab:ks_nrg_n2}, all orbital energies are 
negative. In Table \ref{tab:n2_trans}, the MAE for OEP is $0.34\rm$eV, much lower than
LDA. Since OEP binds all orbitals, it allows many more transitions to be calculated. A common OEP functional is exact exchange 
(or the  KLI approximation\cite{KLI95} to it) which neglects
correlation effects, but these are generally small contributions to the KS orbitals. Using these with ALDA for $f\xc$ 
(which does contain correlation) leads to good transition frequencies as shown in
Table \ref{tab:n2_trans}. Although LDA is sometimes closer to the experimental values for the lower transitions, the value of OEP lies in it 
ability to describe both the higher and lower transisions.

\subsubsection{Napthalene, a small molecule}

\begin{table}
\caption{\label{tab:method_ener} Naphthalene: Effect of ground-state potential 
on the excitations of Table \ref{tab:dft_method_ener}. A ground-state calculation using exact exchange OEP (LHF) is performed and
the excitations are found using a LDA/PBE kernel respectfully. The result is then compared to that found if the LDA/PBE functional 
had been used for both steps.}
\begin{ruledtabular}
\begin{tabular}{c|cccccccc}
Method & 1$\ ^1B_{3u}$ & 1$\ ^1B_{2u}$ & 2$\ ^1A_g$ & 1$\ ^1B_{1g}$
& 2$\ ^1B_{3u}$ & 1$\ ^1A_u$ \\ 
     LSDA  & 4.191 & 4.026 & 5.751 & 4.940 & 5.623 & 5.332 \\
  LHF/LSDA & 4.317 & 4.143 & 5.898 & 5.097 & 5.752 & 5.686 \\
  \hline
     PBE   & 4.193 & 4.031 & 5.753 & 4.957 & 5.622 & 5.141 \\
  LHF/PBE  & 4.295 & 4.121 & 5.876 & 5.091 & 5.741 & 5.693 \\
  \hline
     ``best". &  4.0 & 4.5 & 5.5 & 5.5 &5.5 & 5.7\\
\end{tabular}
\end{ruledtabular}
\end{table}

Returning to our benchmark case of Naphthalene, using more
accurate LHF exchange-only potentials from Sec. \ref{s:gsappr} together with an LSDA or PBE
kernel produces excitation energies in between the GGA and the hybrid
results, except for the 1$\ ^1A_u$ Rydberg state, whose excitation
energy is significantly improved. Whether the LSDA kernel or the PBE GGA
kernel is used together with an LHF potential does not change the results
significantly.

The 1$\ ^1B_{1g}$ and especially the 1$\ ^1A_{u}$ states are diffuse,
and it is not surprising that their excitation energy is considerably
underestimated in the LSDA and GGA treatment. Using the asymptotically
correct LHF potential corrects the excitation energy of the 1$\
^1A_{u}$, which is a pure one-particle excitation out of the $1a_u$
valence into the $10 a_g$ Rydberg orbital; the latter may be viewed as a
$3s$ orbital of the C$_{10}$H$_{8}^+$ ion. On the other hand, a
strong mixture of valence and Rydberg excitations occurs in 1$\
^1B_{1g}$. The LHF potential improves the GGA results only marginally
here, suggesting that more accurate XC kernels are necessary to
properly account for valence-Rydberg mixing.

\subsection{Analyzing the influence of the XC kernel}

In this section, we discuss the importance of the XC kernel in TDDFT
calculations.  As mentioned earlier, the kernels used in practical TDDFT are
local or semi-local in both space and time. Even hybrids are largely semi-local, 
as they only mix in $20-25\%$ exact exchange.

\begin{figure}[htb]
\unitlength1cm
\begin{picture}(12,7.5)
\put(-6.5,-4){\makebox(12,7){
\includegraphics{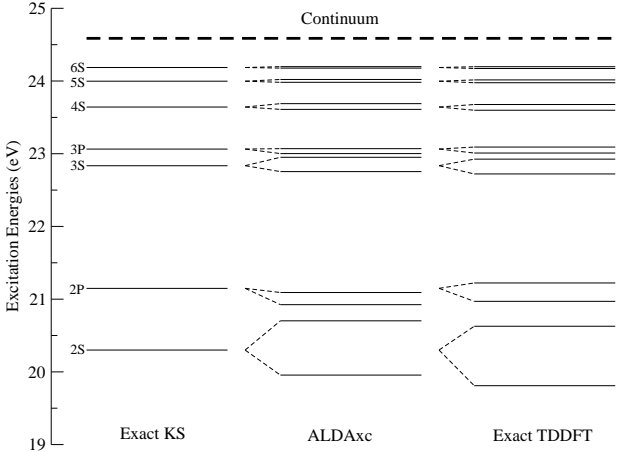}
}}
\end{picture}
\caption{The spectrum of Helium calculated using the ALDA XC kernel\cite{PGB00} with the exact KS orbitals.}
\label{f:He_ALDA}
\end{figure}

In realistic calculations, both the ground-state XC potential and
TDDFT XC kernel are approximated.
A simple way to separate the error in the XC kernel is to look at
a test case where the exact KS potential is known.  Figure \ref{f:He_ALDA} shows the
spectrum of He using the exact KS potential, but with the ALDA XC kernel.
It does rather well\cite{PGB00} (very well, as shall see later in section \ref{s:atoms}, when we examine atoms
in more detail).  Very similar results are obtained with standard GGA's.

\begin{figure}[htb]
\unitlength1cm
\begin{picture}(12,7.5)
\put(-6.5,-4){\makebox(12,7){
\includegraphics{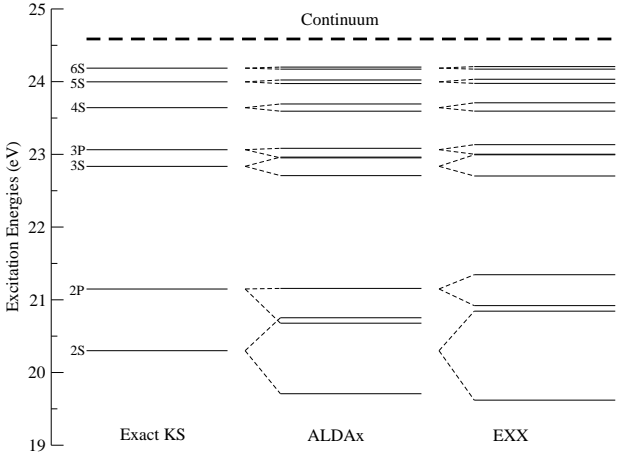}
}}
\end{picture}
\caption{The spectrum of Helium calculated using the \mbox{ALDAx} kernel and the exact exchange kernel\cite{PGB00}. Again the exact KS orbitals were used. The importance of non-locality for the XC kernel can be seen as the exchange part of ALDA gives a noticeable error compared to the exchange part of the true functional (the AEXX kernel for He).}
\label{f:He_ALDAx}
\end{figure}

The errors in such approximate kernels come from the locality in space and time.
We can test one of these separately for the He atom, by studying the exchange
limit for the XC kernel.  For two spin-unpolarized electrons, $f\x=-1/2|\br-\br'|$, i.e., it exactly
cancels half the Hartree term.  Most importantly, it is frequency-independent,
so that there is no memory, i.e., the adiabatic approximation is exact. In Fig. \ref{f:He_ALDAx}, we compare
ALDAx, i.e., the ALDA for just exchange, to the exact exchange result for He.
Clearly, ALDA makes noticeable errors relative to exact exchange, showing that
non-locality in space can be important.

Thus the hybrid functionals, by virtue of mixing some fraction of exact exchange with GGA, 
will have only slightly different potentials ( mostly in the asymptotic region),
but noticeably different kernels.

\subsection{Errors in potential vs. kernel}

In this section, we examine the relative importance of the
potential and kernel errors.  It has long been believed that fixing the
defects in the potential, especially its asymptotic behavior, has been
the major challenge to improving TDDFT results\cite{TH98,WY03,CS00}.  We argue here that this
is overly simplistic, and is due to tests being carried out on atoms
and small molecules.   In large molecules, where the interest is in the
many low-lying transitions, the potential can be sufficiently accurate, while the kernel
may play a larger role.

In fact, our analysis of the general failure of TDDFT
in underestimating the $^1L_a$
transitions in PAH's  sheds some light on its origin.
Using the self-interaction free LHF potential does not
cure this problem, as is obvious from Tab. \ref{tab:method_ener}. To
the best of our knowledge, the cause of this shortcoming of TDDFT is
not well understood. We note, however, that the same incorrect
ordering of $^1L_a$ and $^1L_b$ occurs in the CIS approximation, which
is self-interaction free.  The analysis here shows that this is
a failure of our approximations to the  XC kernel rather than
to the ground-state potential.

\subsection{Understanding linear response TDDFT}
\label{s:under}
Several simple methods have evolved for qualitatively understanding
TDDFT results. The most basic is the single-pole approximation (SPA), which originated\cite{PGG96}
in including only one pole of the response function.  The easiest way
to see this here is to truncate Eq. (\ref{Casida}) to a 1$\times$1 matrix,
yielding an (often excellent) approximation to the change in transition
frequency away from its KS value\cite{VOC99,VOC02}:
\ben
\omega^{2} \approx w_{q\sigma}^{2} + 2\omega_{q\sigma}K_{q\sigma q\sigma} \ \ (\mbox{SPA}),
\een
(The original SPA was on the unsymmetric system yielding $\omega \approx w_{q\sigma} + K_{q\sigma q\sigma}$, which 
for a spin-saturated system becomes $\omega \approx \omega_q + 2K_{qq}$ \cite{PGB00}).
This can also provide a quick and dirty estimate, since only KS transitions
and one integral over $f\xc$ are needed.  While it allows an estimate
of the shift of transitions from their KS energy eigenvalue differences,
it says nothing about oscillator strengths, which are unchanged in SPA
from their
KS values.  In fact, a careful analysis of the TDDFT equation shows that
oscillator strengths are particularly sensitive to even small off-diagonal
matrix elements, whereas transition frequencies are less so\cite{AGB03}.

A more advanced analysis is the double pole approximation\cite{AGB06} (DPA), which 
applies when two transition
are strongly coupled to one another, but not strongly
to the rest of the transitions.   Then one can show explicitly the
very strong effect that off-diagonal elements have on oscillator
strengths, showing that sometimes an entire peak can have almost
no contribution.  One also sees pole-repulsion in the positions of
the transitions, a phenomenon again missing from SPA.

The DPA was used recently and very
successfully to explain X-ray edge spectroscopy results for
$3d$-transition metal solids as one moves across the periodic
table\cite{SGAS05}.  These transitions form a perfect test case for DPA, as
the only difference between them is caused by the spin-orbit
splitting (several eV) of the 2p$^{1/2}$ and 2p$^{3/2}$ levels.
In a ground-state KS calculation, this leads to a 2:1 branching
ratio for the two peaks, based simply on degenearcy, as all
matrix elements are identical for the two transitions.
Experimentally, while this ratio is observed for Fe, 
large deviations occur for other elements.

These deviations
could be seen in full TDDFT calculations, and were attributed to
strong core-hole correlations.  The SPA, while it nicely accounts for the
shifts in transition frequencies relative to bare KS transitions,
but yields only the ideal 2:1 branching ratio. However, the DPA model gives a much simpler, and  more benign
interpretation.  The sensitivity of oscillator strengths to
off-diagonal matrix elements means that, even when the off-diagonal
elements are much smaller than diagonal elements (of order 1 eV),
they cause rotations in the 2-level space, and greatly alter the
branching ratio. Thus a KS branching ratio occurs even with strong diagonal `correlation', so long as off-diagonal XC contributions are truly negligible. But even small off-diagonal `correlation', can lead to large deviations from KS branching ratios.

\begin{table}
\caption{\label{tab:DPA} Transition frequencies and oscillator strengths (O.S) calculated using the  double pole approximation (DPA) for the lowest $^{1}B_{3u}$ transitions in Naphthalene. The PBE functional was used with a aug-TZVP basis set on top of a PBE/TZVP/RI ground state structure}
\begin{ruledtabular}
\begin{tabular}{c||cc|cc|cc}
	& \multicolumn{2}{c|}{KS} & \multicolumn{2}{c|}{DPA} & \multicolumn{2}{c}{Full TDDFT} \\
	& $\omega$ &  O.S & $\omega$ & O.S & $\omega$ & O.S \\
\hline	
$^{1}B_{3u}$ & 4.117 & (1.02) & 4.245 & (0.001)  & 4.191 & (0) \\
$^{2}B_{3u}$ & 4.131 & (1.00) & 6.748 & (2.02)	 & 5.633 & (1.14)\\

\end{tabular}
\end{ruledtabular}
\end{table}

\begin{figure}
  \centering
  \includegraphics[width=5cm]{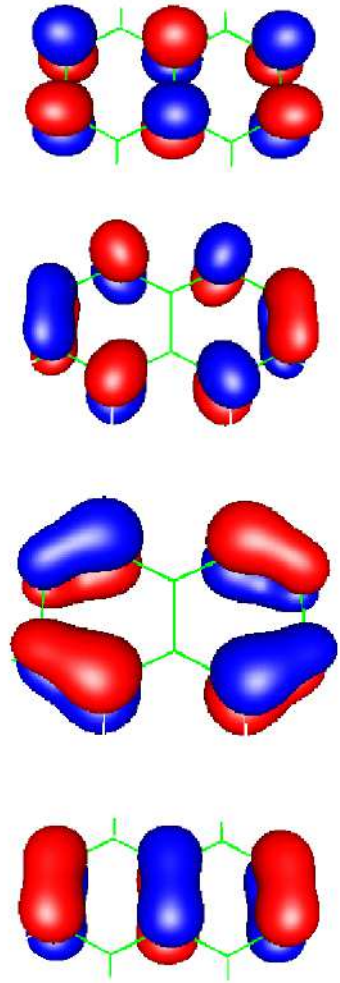}
  \caption{The four orbitals involved in the first two $\
    ^1B_{3u}$ (contour value $\pm$ 0.07 a.u.). The PBE functionals and
    an aug-TZVP basis set were used.} 
  \label{fig:naph_orb}
\end{figure}

We can use DPA  to understand the lowest $\ ^1B_{3u}$
transitions in our naphthalene case.
In Table \ref{tab:DPA}, we list the TDDFT matrix elements for the PBE calculation
for the 
two nearly degenerate KS transitions, $1a_u
\rightarrow 2b_{3g}$ and $2b_{1u} \rightarrow 2b_{2g}$, along
with their corresponding KS transition frequencies.
Contour plots
of the four orbitals involved are shown in Fig. \ref{fig:naph_orb}.
We note first that these two KS transitions are essentially degenerate,
so that there is no way to treat them within SPA.  The degeneracy
is lifted by the off-diagonal elements, which cause the transitions
to repel each other, and strongly rotate the oscillator strength between
the levels, removing almost all the oscillator strength from the lower
peak\cite{AGB06}.  The DPA yields almost the correct frequency and oscillator strength
(i.e., none) for the lower transition, but the higher one is overestimated, with
too much oscillator strength.  This must be due to coupling to other
higher transitions.  In the DPA, in fact the higher transition lands right
on top of the third transition, so strong coupling occurs there too.
This example illustrates (i), that solution of the full TDDFT equations is typically necessary for large 
molecules which have many coupled transitions, but also (ii), that simple models can aid the interpretation of such results.
All of which shows that, while models developed for well-separated transitions
might provide some insight for specific transitions in large molecules,
the number and density of transitions make such models only semi-quantitative
at best.

\section{Atoms as a test case}
\label{s:atoms}

In this section, we look more closely at how well TDDFT performs
for a few noble gas atoms.  As explained above, this is far from 
representative of its behavior for large molecules, but 
this does allow careful study of the
electronic spectra without other complications.  Most importantly,
for the He, Be, and Ne atoms, we have essentially exact ground-state
KS potentials from Umrigar and coworkers\cite{UG93,UG94}.   This allows us to dissect the
sources of error in TDDFT calculations.

\subsection{Quantum defect}
\label{s:qd}

\begin{figure}
\includegraphics[width=3.3in]{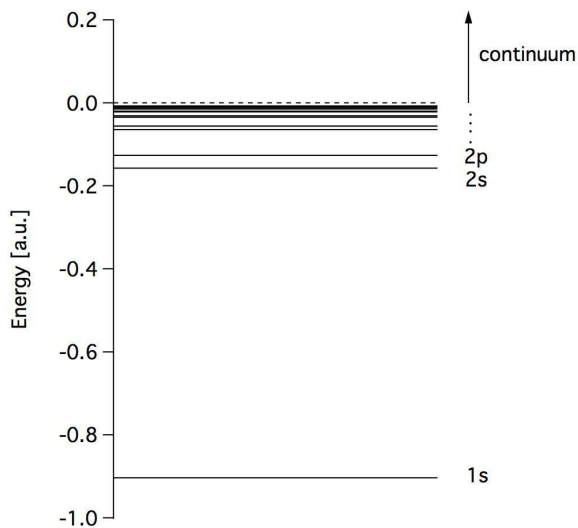}
\caption{\label{fig:helevels} Singlet energy level diagram for the helium atom. The Rydberg series of transition frequencies 
clustered below the ionization threshold can be seen. The frequencies cluster together, making it difficult to assess the 
quality of the TDDFT calculated spectra. As discussed in the text, the quantum defect is preferable for this purpose.}
\end{figure}

In Fig.~\ref{fig:helevels} we show the KS orbital energy level diagram of
the helium atom.  The zero is set at the onset of the continuum and is marked
with a dotted
line. For closed shell atoms and
for any spherical one-electron potential that decays as $-1/r$ at
large distances, the bound-state transitions form a Rydberg series
with frequencies:
\begin{equation}
\omega_{nl}=I-\frac{1}{2(n-\mu_{nl})^2}
\label{eq:qd}
\end{equation}
where $I$ is the
ionization potential, and
$\mu_{nl}$ is called the quantum defect.
Quantum defect theory was
developed by Ham \cite{H55} and Seaton
\cite{S58} before even the Hohenberg-Kohn theorem\cite{HK64}.

The great value of the quantum defect is its
ability to capture all the information about the entire
Rydberg series of transitions in a single slowly-varying
function, the quantum defect as a function of energy, $\mu_l(E=\omega-I)$,
which can often be fit by a straight line or parabola. In Table \ref{tab:helium}, we report 
extremely accurate results from
wavefunction calculations for the helium atom. We show singlet and
triplet values that have been obtained by Drake~\cite{D96}. We also
give results from the exact ground-state KS potential shown in Fig \ref{f:He_ext_ks}~\cite{UG94}.
For each column, on the left are the
transition frequencies, while on the right are the corresponding
quantum defects.  Note how small the differences between transitions
become as one climbs up the ladder, and yet the quantum defect remains
finite and converges to a definite value.

\begin{table}
\caption{\label{tab:helium} Transition energies ($\omega$) and quantum defects (QD) for He atom $s$-Rydberg
series[a.u.].
The ionization energy is $0.90372$ a.u.}
\begin{ruledtabular}
\begin{tabular}{c||cc|cc|cc}
{Transition} & \multicolumn{2}{c|}{Singlet \footnotemark[1]} &
\multicolumn{2}{c|}{Triplet \footnotemark[1]} & \multicolumn{2}{c}{KS\footnotemark[2]}\\
& $\omega$ & QD & $\omega$ & QD& $\omega$ & QD\\
\hline
$1s\rightarrow 2s$   & 0.7578  & 0.1493	& 0.7285 & 0.3108	& 0.7459 & 0.2196 \\
$1s\rightarrow 3s$   & 0.8425  & 0.1434	& 0.8350 & 0.3020	& 0.8391 & 0.2169\\
  $1s\rightarrow 4s$ & 0.8701  & 0.1417	& 0.8672 & 0.2994	& 0.8688 & 0.2149 \\
  $1s\rightarrow 5s$ & 0.8825  & 0.1409	& 0.8883 & 0.2984	& 0.8818 & 0.2146 \\
  $1s\rightarrow 6s$ & 0.8892  & 0.1405	& 0.8926 & 0.2978	& 0.8888 & 0.2144 \\
  $1s\rightarrow 7s$ & 0.8931  & 0.1403	& 0.8926 & 0.2975	& 0.8929 & 0.2143
\end{tabular}
\end{ruledtabular}
\footnotetext[1]{Accurate non-relativistic calculations from Ref.~\cite{D96}.}
\footnotetext[2]{The differences between the KS eigenvalues obtained
with the exact potential from Ref.~\cite{UG94}.}
\end{table}

\begin{figure}
\includegraphics[width=3.3in]{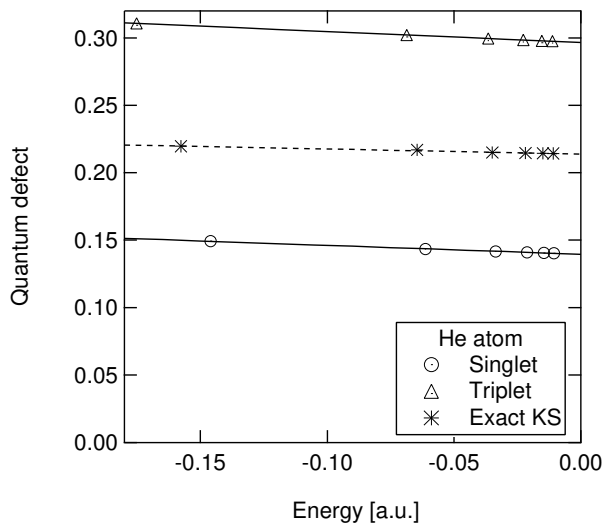}
\caption{\label{fig:exactvse} The exact \emph{s} KS quantum defect and the exact singlet and triplet quantum defects of He and their parabolic fits. The quantum defect may clearly be described as a smooth function of energy, in this case, a linear fit. Thus knowing the quantum defect for a few transitions allows us to find it for all transitions, and hence their frequencies.}
\end{figure}
All the information of the levels of Fig. \ref{fig:helevels} and of Table \ref{tab:helium} is contained
in
Fig.~\ref{fig:exactvse}.
This clearly illustrates that the
quantum defect is a smooth function of energy, and is well
approximated (in these cases) as a straight line. The quantum defect is thus an extremely compact and sensitive test of
approximations to transition frequencies.
Any
approximate ground-state KS potential suggested for use in TDDFT
should have its quantum defect compared with the exact KS quantum defect,
while any approximate XC-kernel should produce accurate corrections
to the ground-state KS quantum defect, on the scale of Fig. \ref{fig:exactvse}.

To demonstrate the power of this analysis, we test two common
approximations to the ground-state potential, both of which produce
asymptotically correct potentials.
These are exact
exchange~\cite{TS76} (see Sec \ref{s:gsappr}) and LB94~\cite{LB94}. Exact exchange
calculations are more demanding than traditional DFT calculations, but
are becoming popular because of the high quality of the
potential~\cite{G99,IHB99}. On the other hand, LB94 provides an
asymptotically correct potential at little extra cost beyond
traditional DFT~\cite{CS00,GGGB01,WAY03}.
 \begin{figure}
\includegraphics[width=3.3in]{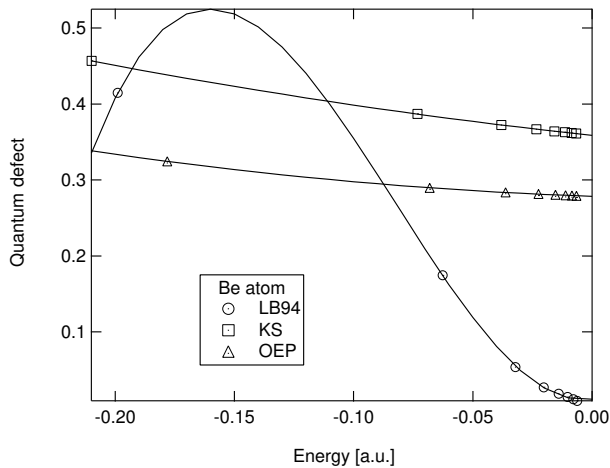}
\caption{\label{fig:lb94fit} The Be $p$ quantum defect of LB94, exact exchange (OEP),
and KS, and their best fits. While both functionals give the correct asymptotic behavior of the KS potential, by calculating the 
quantum defect, we can learn more about their performance.}
\end{figure}
In Fig.~\ref{fig:lb94fit} we show the $p$ Be quantum defect obtained
with LB94, OEP, and exact KS potentials.
Fig.~\ref{fig:lb94fit} immediately shows the high quality of the exact exchange
potential. The quantum defect curve is almost identical to the exact
one, apart from being offset by about 0.1 . On the
other hand, the quantum defect of LB94 was poor for
all cases studied\cite{FB06,Fc06}. This shows that just having a potential that is
asymptotically correct is not enough to get a good quantum defect.

\subsection{Testing TDDFT}
\label{s:qdDFT}

\begin{figure}
  \centering
  \includegraphics[width=3.6in]{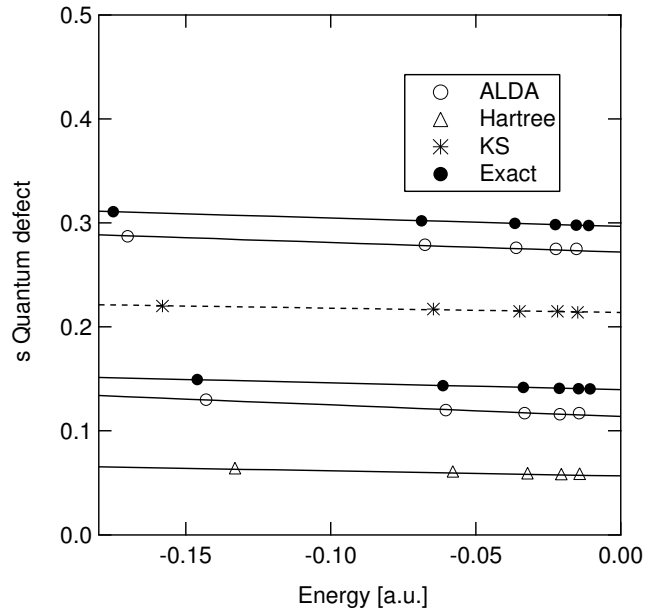}
  \caption{The corrections due to using the Hartree or ALDA kernel on the exact KS \emph{s} quantum defect of He. Using 
  the Hartree kernel only effects the singlet values, shifting them too low. If a good XC kernel is then used, it should move 
  both the triplet and singlet quantum defects from the Hartree kernel towards the exact ones\cite{Fc06}. In this case, ALDA does a good job 
  and is performing well.} 
  \label{f:HeQD}
\end{figure}
To see how well TDDFT really does, we plot quantum defects
for atoms.  We take the He atom as our prototype, as usual in this section.
In Fig. \ref{f:HeQD}, we plot first the KS quantum defect and the exact singlet and triplet lines, as before in Fig. \ref{fig:exactvse}.
Then we consider the
Hartree approximation.  This is equivalent to setting the XC kernel
to zero.  This changes the postion of the singlet curve, but leaves
the triplet unchanged from its KS value, because the direct term
includes no spin-flipping.  It definitely improves over the KS for the
singlet.  Lastly, we include ALDA XC corrections.  Only if these
significantly improve over the Hartree curves can we say TDDFT
is really working here.  Clearly it does, reducing the Hartree error
enormously.

These results are also typical of He P transitions, and Be S and P transitions.
For reasons as yet unclear, the $s\to d$ transitions fail badly for both
these systems\cite{FB06,FBb06}.

\subsection{Saving standard functionals}
\label{s:trunc}

We have a problem with the  incorrect long-range behavior
of the potential from standard {\em density} functionals only when Rydberg excitations are 
needed.
But it would be unsatisfactory to have to perform a completely different
type of calculation, eg OEP, in order to include such excitations
when desired, especially if the cost of that calculation is
significantly greater.

\begin{figure}
\includegraphics[scale=0.9]{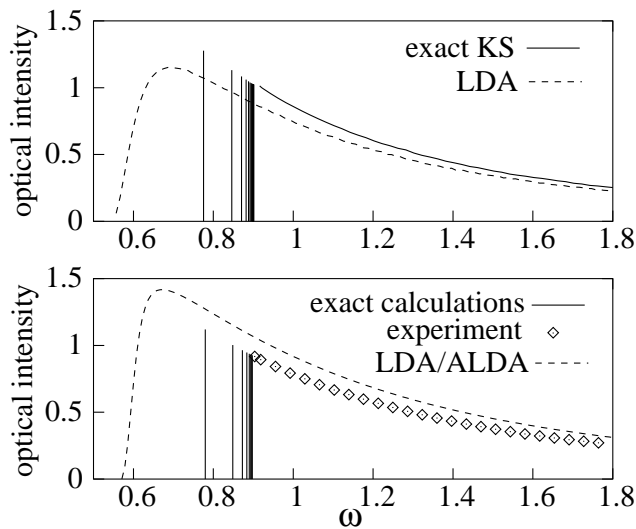}
\caption{He atom: The top panel shows the bare exact KS and LDA spectra, and the lower panel shows the TDDFT corrected spectra,  LDA/ALDA results are from
~\cite{SDG01} but unshifted; the exact 
calculations are from \cite{KH84},
multiplied by the density of states factor (see text),
and the experimental results are from \cite{SHYH94}}.
\label{f:He}
\end{figure}

However, it is possible, with some thought and care, and using
quantum defect theory, to extract the Rydberg series from the 
short-ranged LDA potential.  To see this, consider Fig. \ref{f:He}, which shows both the bare KS response and the TDDFT
corrected response of the He atom.
The $\delta$-function absorptions at the discrete transitions have
been replaced by straightlines, whose height represents the oscillator
strength of the absorption, multiplied by the appropriate density
of states\cite{F98}.
In the top panel, just the KS transitions are shown, for both the
KS potential and the LDA potential of Fig \ref{f:LDA_vs_exKS} from section \ref{s:gsappr}.
The exact curve has a Rydberg series converging to 0.903, the exact
ionization threshold for He.  The LDA curve, on the other hand, has
a threshold at just below 0.6.  But clearly its optical absorption
mimics that of the exact system, even in the Rydberg series region, and is accurate to about 20\%.
The TDDFT ALDA corrections
are small, and overcorrect the bare LDA results, but clearly are
consistent with our observations for the bare spectra.

Why is this the case? Is this a coincidence?  Returning to Fig. \ref{f:LDA_vs_exKS} of the introduction, we notice
that the LDA (or GGA) potential runs almost exactly parallel to the
true potential for $r \lesssim 2$, i.e., where the density is.
Thus the scattering orbitals of the LDA potential, with transition
energies between 0.6 and 0.9, almost exactly match the Rydberg
orbitals of the exact KS potential with the same energy. When carefully defined, i.e., phase space 
factors for the continuum relative to bound states, the oscillator strength  is about the same. This is no 
coincidence but, due to the lack of derivative discontinuity of LDA, its potential differs from 
the exact one by roughly a constant.

The `fruitfly' of TDDFT benchmarks is the $\pi\to\pi^*$ transition
in benzene.  This occurs at about 5 eV in a ground-state LDA calculation,
and ALDA shifts it correctly to about 7 eV\cite{VOC02}.
Unfortunately, this is
in the LDA continuum, which starts at about 6.5 eV! 
This is an example of the same general phenomenon, where LDA has
pushed some oscillator strength into the continuum, but its overall
contribution remains about right.

We can go one step further, and even deduce the energies of individual
transitions.
While the existence of a quantum defect requires a long-ranged
potential, its value is determined by the phase-shift caused by
the deviation from $-1/r$ in the interior of the atom. 
The {\em quantum defect extractor} (QDE)\cite{WBb05},
 is a formula for extracting the effective quantum defect from
a scattering orbital of a short-ranged KS potential, such as that
of LDA.   The QDE is:
\ben \frac{d\ln
\phi}{dr}=\frac{1}{n^*}-\frac{n^*}{r}-\frac{1}{r}\frac{U(-n^*;2;2r/n^*)}{U(1-n^*;2;2r/n^*)}
\label{ld}\een
Here $k={\sqrt{2|E|}}$ is
written as $k=(n^*)^{-1}$, with $n^*=(n-\mu_n)$, where $n$
numbers the bound state, and $\mu_n$ is the quantum defect; $U$ is the
confluent hypergeometric function
\cite{AS72}.
If the extractor is applied to an orbital of a long-ranged potential,
it rapidly approaches its quantum defect.

\begin{figure}
  \centering
  \includegraphics[width=3.3in]{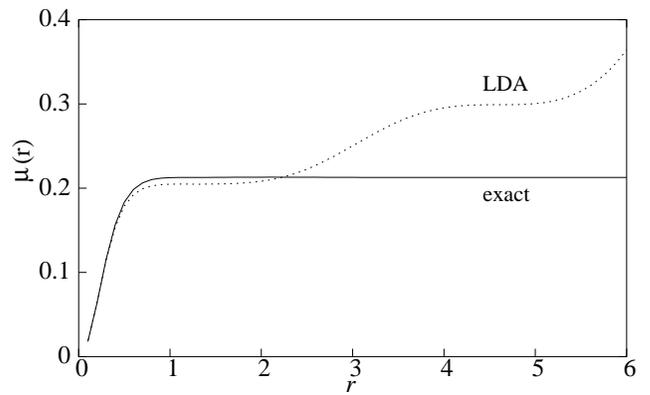}
  \caption{He atom: solution of Eq.(\ref{ld}) for $\mu$ as a function of $r$; The $n=20$ orbital was used for the exact case, and the scattering orbital or energy $E=I+\epsilon_{1s}^{LDA}$ was used for the LDA.} 
  \label{f:He_LDA_and_exact_qd}
\end{figure}

In Fig \ref{f:He_LDA_and_exact_qd}, we plot the results of the QDE for the He atom, applied to both
the exact KS potential and the LDA potential.
The LDA potential runs almost parallel to the exact one in the region
$1<r<2$ (where $\mu_{\infty}$ can already be extracted accurately), and
orbitals corresponding to the same {\em frequency} (exact and LDA) are
therefore very close in that region. In
the spirit of Ref.\cite{WMB03}, we compare the exact
energy-normalized 20s orbital (which is essentially identical to the zero-energy state
in the region $0<r<6$) and the LDA orbital of energy
$I+\epsilon_{1s}^{LDA}=0.904-0.571=0.333$. Notice how good the LDA
orbital is in the region $1<r<2$. We show in
Fig.\ref{f:He_LDA_and_exact_qd} the solution of Eq.(\ref{ld}) when
this scattering LDA orbital is employed. Clearly, the plateau of the
LDA curve in the $1<r<2$ region is an accurate estimate of the quantum
defect. The value of $\mu$ on this plateau is 0.205, an
underestimation of less than 4\% with respect to the exact value.

Thus, given the ionization potential of the system, LDA gives a very
accurate prediction of the asymptotic quantum defect. The ionization
potential is needed to choose the appropriate LDA scattering orbital,
but the results are not terribly sensitive to it. We repeated the same
procedure with the LDA ionization potential (defined as
$E_{\rm LDA}$(He)$-E_{\rm LDA}$(He$^+$)=0.974) instead of the exact
one, and found $\mu_{\infty}^{\rm LDA}=0.216$,
overestimating the exact $\mu_{\infty}$ by just 1\%.

\subsection{Electron scattering}
\label{s:scatt}

Lastly in this section, we mention recent progress in developing a theory for
low-energy electron scattering from molecules.  This was one of the original
motivations for developing TDDFT.   One approach would be to
evolve a wavepacket using the TDKS equations, but a more direct
approach has been developed\cite{WMB05,WB06},
in terms of the response function $\chi$
of the $N+1$ electron system (assuming it is bound).

This uses similar
technology to the discrete transition case.  Initial results for
the simplest case, electron scattering from He$^+$, suggest
a level of accuracy comparable to bound-bound transitions, at least
for low energies (the most difficult case for traditional methods,
due to bound-free correlation\cite{N00}).
TDDFT, using the exact ground-state potential
and ALDA, produces more accurate answers than static exchange\cite{WMB05},
a traditional low cost method that is used for larger 
molecules\cite{TG05,TGb06}.

However, that TDDFT method fails when applied to electron scattering from
Hydrogen, the true prototype, as the approximate solution of the
TDDFT equations (very similar to the single pole approximation of
Sec \ref{s:under}) fails, due to stronger correlations.  To overcome this,
a much simpler method has been developed, that uses an old
scattering trick\cite{F05} to deduce scattering phase shifts from
bound-state energies when the system is placed in a box, yielding
excellent results for a very demanding case.

\section{Beyond standard functionals}
\label{s:beyond}

We have surveyed and illustrated some of the many present successful
applications of TDDFT in the previous section.  In these
applications, standard approximations (local, gradient-corrected,
and hybrid, see sec. \ref{s:gsappr}) are used both for the ground-state calculation
and the excitations, via the adiabatic approximation (sec. \ref{s:tdappr}).
In this section, we survey several important areas in which this
approach has been found to {\em fail}, and what might be done about it.

The errors are due to locality in both space and time, and these
are intimately related.  In fact, all 
memory effects, i.e., dependence on the history of the density\cite{HMB02},
implying a frequency-dependence in the XC kernel,
can be subsumed into an initial-state dependence\cite{MBW02},
but probably not vice-versa.  Several groups are attempting to
build such effects into new approximate functionals
\cite{UB04,UT06,U06,V01,TV06,KB04,KB05,KB06,T05,Tb05}, but none have
shown universal applicability yet.

The failure of the adiabatic approximation
is most noticeable when higher-order excitations are considered,
and found to be missing in the usual linear response treatment\cite{C96}.  The failure of
the local approximation in space is seen when TDDFT is applied
to extended systems, e.g., polymers or solids.  Local approximations
yield short-ranged XC kernels, which become irrelevant compared
to Hartree contributions in the long-wavelength limit.
The Coulomb repulsion between electrons generally requires long-ranged
(ie $1/r$) exchange effects when long-wavelength response is being
calculated.

Thus several approaches have been developed and applied in places
where the standard formulation has failed.
These approaches fall into two distinct categories.  On the one hand,
where approximations  that are local in the {\em density} fail,
approximations that are local (or semi-local) in the
{\em current-density} might work.  In fact, for TDDFT, the gradient
expansion, producing the leading corrections to ALDA, {\em only}
works if the current is the basic variable\cite{V06}.  Using the gradient
expansion itself is called the Vignale-Kohn (VK) approximation\cite{VK96,VUC97}, and
it has been tried on a variety of problems.

The alternative approach is to construct orbital-dependent
approximations with explicit frequency-dependence\cite{MZCB04,MSR06}.  This can
work well for specific cases, but it is then hard to see
how to construct general density functional approximations
from these examples.  More importantly, solution of the OEP equations
is typically far more expensive than the simple KS equations,
making OEP impractical for large molecules.

\subsection{Double excitations}
\label{s:double}

As first pointed out by Casida\cite{C96}, double excitations appear
to be missing from TDDFT linear response, within any adiabatic approximation.
Experience\cite{TAHR99,TH00} shows that, like in naphthalene,
sometimes adiabatic TDDFT will produce a single excitation in about the
right region, in place of two lines, where a double has mixed strongly
with a single.

In fact, when a double excitation lies close to a single excitation,
elementary quantum mechanics shows that $f\xc$ must have a
strong frequency dependence
\cite{MZCB04}.
Recently, post-adiabatic TDDFT methodologies have been
developed\cite{MZCB04,ZB04,C05} for
including a double excitation when it is close to
an optically-active single excitation, and works
well for small dienes\cite{CZMB04,MZCB04}.
It might be hoped that, by
going beyond linear response, non-trivial double excitations would be
naturally included in, e.g.,
TDLDA, but it has recently been proven
that, in the higher-order response in TDLDA, the double excitations
occur simply at the sum of single-excitations\cite{TC03}.
Thus we do not currently
know how best to approximate these excitations.  This problem is particularly
severe for quantum wells, where the external potential is parabolic,
leading to multiple near degeneracies between levels of excitation\cite{ZB04}.

Returning to our naphthalene example, based on a HF reference, the 2$\ ^1A_g$ state has, according to the RICC2
results, a considerable admixture of double excitations. This is
consistent with the fact that the CIS method yields an excitation
energy that is too high by 1.5 eV compared to experiment. The TDDFT
results are much closer, yet too high by several tenths of eV.

\subsection{Polymers}
\label{s:poly}

\begin{figure}[htb]
\unitlength1cm
\begin{picture}(12,6.2)
\put(-6.4,-4){\makebox(12,7){
\includegraphics{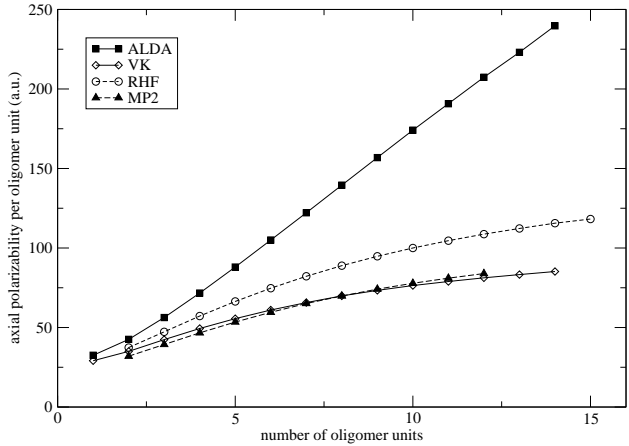}
}}

\end{picture}
\caption{ALDA and VK static axial polarizability of polyacetylene
compared with RHF and MP$2$ results from Refs. \cite{FBLB02,FBLB03,F06}. ALDA severely overestimates the polarizability compared to 
the accurate MP2 calculation. Hartree-Fock is also incorrect. However using the VK functional gives almost exact agreement, at least 
in this case.}
\label{f:poly_pol}
\end{figure}
An early triumph of the VK functional was the static polarizabilities of
long-chain conjugated polymers.  These polarizabilites are
greatly underestimated by LDA or GGA, with the error growing rapidly
with the number of units\cite{GSGB99}.  On the other hand, HF does rather well, and
does not overpolarize.  The VK correction to LDA yields excellent results
in many (but not all) cases, showing that a current-dependent functional
can correct the over-polarization problem.  Naturally, orbital-dependent
functionals also account for this effect\cite{KKP04}, but at much higher computational cost.

\subsection{Solids}
\label{s:solids}

Again, in trying to use TDDFT to calculate the optical response
of insulators, local approximations has been shown to fail badly.
Most noticeably, they do not describe excitonic effects\cite{KSG03}, or the
exciton spectrum within the band gap.  On top of this, the gap is
usually much smaller than experiment, because adiabatic appoximations
cannot change the gap size from its KS value.

One approach is using the VK approximation in TDCDFT.  This has proven
rather successful, although a single empirical factor was needed to
get agreement with experiment\cite{KBS00,KBSb00,BKBL01}.
An alternative is to study the many-body problem\cite{ORR02}, and ask which expressions
must the XC kernel include in order to yield an accurate absorption spectrum\cite{SOR03,MSR03}.
However, the presently available
schemes require an expensive GW calculation in the first place\cite{MR04}.
A recent review can be found in Ref. \cite{MUNR06}.

\subsection{Charge transfer}
\label{s:char}

As is usually the case whenever a method is shown to work well,
it starts being applied to many cases, and specific failures
appear.  Charge transfer excitations are of great importance
in photochemistry, especially of biological systems, but
many workers have now found abysmal results with TDDFT for
these cases.

This can be understood from the fact that TDDFT is a linear response
theory.  When an excitation moves charge from one area in a molecule
to another, both ends will relax.  In fact, charge transfer between
molecules can be well-approximated by ground-state density functional
calculations of the total energies of the species involved.  But
TDDFT must deduce the correct transitions by infinitesimal perturbations
around the ground-state, without an relaxation.  Thus it seems a
poor problem to tackle with linear response.  Many researchers
are studying this problem, to understand it and find practical
solutions around it\cite{Mc05,MT05,HG06,JC04,CGGG00}.

\section{Other topics}
\label{s:other}

In this chapter, we discuss several topics of specialized interest, where
TDDFT is being applied and developed in ways other than simple extraction
of excitations from linear response.

In the first of these, we show how TDDFT can be used to construct
entirely new approximations to the ground-state XC energy.  This
method is particularly useful for capturing the long-range fluctuations
that produce dispersion forces between molecules, which are notoriously
absent from most ground-state approximations.

In the second, we briefly survey strong field applications, in which
TDDFT is being used to model atoms and molecules in strong laser fields.
We find that it works well and easily for some properties, but less
so for others.

In the last, we discuss the more recent, hot area of molecular electronics.
Here, many workers are using ground-state DFT to calculate transport
characteristics, but a more careful formulation can be done only within
(and beyond) TDDFT.  We review recent progress toward a more rigorous
formulation of this problem.

\subsection{Ground-state XC energy}
\label{s:gs}
\begin{figure}
\includegraphics[width=3.3in]{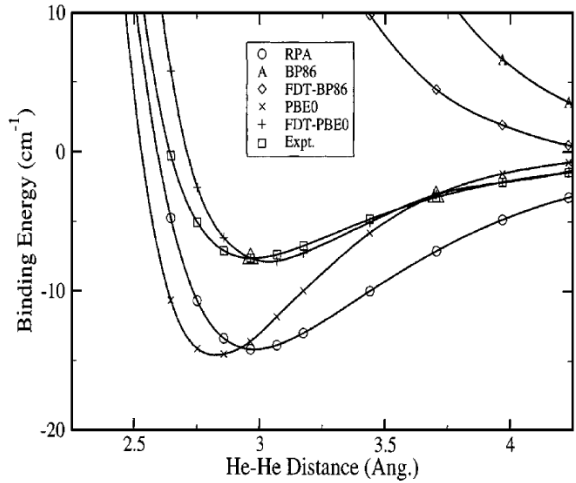}
\caption{\label{f:He_He_vdw} Binding energy for the Helium dimer interacting via Van der Waals (VdW) forces, from Ref. \cite{FV05}.
 Using the fluctuation-dissipation theorem (FDT), new XC energy functionals may be constructed using any ground-state functional. 
 The curves from the standard ground-state functionals BP86\cite{B88,P86} and PBE\cite{PBE96} are given, as well as the FDT- curves with 
 these as input. Clearly the FDT is needed to accurately describe VdW interaction.}
\end{figure}

TDDFT offers a method\cite{LGP00,ALL96,FNGB05,DRSL04,FV05} to find more sophisticated ground-state approximate energy functional using the 
frequency-dependent response function. Below we introduce the basic formula and discuss some of the exciting systems this method is being 
used to study.\\

This procedure uses the adiabatic connection fluctuation-dissipation formula:
\ben
\label{Exc}
E\xc[\n_0] = \half \int_0^1\; d\lambda\;\int d^3r\; \int d^3r'\;\frac{P\l(r,r')}{|\br-\br'|}
\een
where the pair density is
\bea
P\l(r,r') &=& - \left(\dssum\int_0^\infty\frac{d\omega}{\pi}\chi\l\dsig[\n_0](\br\br';i\omega)\right) \nonumber \\
& & \hspace{0.2cm} - n_0(\br) \delta^{(3)}(\br-\br')\nonumber
\eea
and the coupling-constant $\lambda$ is defined to multiply the
electron-electron repulsion in the Hamiltonian, but the external
potential is adjusted to keep the density fixed\cite{LP75,GL76}. $\chi\l\dsig$ is given by Eq (\ref{Dyson}) with the XC kernel $f\l_{{\sss XC}\sigma\sigma'}$. Any approximation to the XC kernel yields a sophisticated XC energy $E\xc[\n]$.\\

It is interesting that if we set XC effects to zero in conventional DFT, we end up with the highly inaccurate Hartree method of $1928$. However when calculating the linear response, if the XC kernel is zero (i.e. within the random phase approximation), the XC energy calculated using Eq. (\ref{Exc}) still gives useful results.\\

Computationally this procedure is far more demanding than conventional
DFT, but as the above example has shown, even poor approximations to
the XC kernel can still lead to good results. Using this method to
find the XC energy has the ability to capture effects such as
dynamical correlation or Van der Waal interactions, which are missing
from conventional ground-state DFT approximations, and are thought to
be important in biological systems.  

In particular, the coefficient in the decay of the energy between
two such pieces ($C_6$ in $E\to -C_6/R^6$, where $R$ is their separation)
can be accurately (within
about 20\%) evaluated using a local approximation to the frequency-dependent
polarizability\cite{ALL96,HALL96,OGSB97,KMM98}.
In Fig. \ref{f:He_He_vdw}, the binding energy curve for two Helium
atoms interacting via Van der Waals is shown. Using the
fluctuation-dissipation formula,  Eq. (\ref{Exc}), and the PBE$0$ XC
kernel clearly gives more accurate results than semi-local
functionals. Recently, the 
frequency integral in Eq. (\ref{Exc})
has been performed explicitly but approximately, yielding an
explicit non-local density functional\cite{ALL96,HALL96,AHRA97,RLLD00,RDJS03,DRSL04,KSL06,PDL06,TPL06} applicable at all
separations.
TDDFT response functions have also been used in the framework of
symmetry-adapted perturbation theory to generate accurate binding
energy curves of Van der Waals molecules \cite{MJS03}.\\

One can go the other way, and try using Eq. (\ref{Exc}) for all
bond lengths\cite{F01,FG02}.In fact,  Eq (\ref{Exc}) provides a KS 
density functional that allows bond-breaking without artifical symmetry
breaking\cite{FNGB05}.
In the paradigm case of the H$_2$ molecule, the binding energy
curve has no Coulson-Fischer point, and the dissociation occurs correctly
to two isolated H atoms.  Unfortunately, simple approximations, while
yielding correct results near equilibrium and at infinity, produce an
unphysical repulsion at large but finite separations.
This can be traced back\cite{FNGB05} to the
lack of double excitations in any adiabatic $f\xc$. Study of the convergence of $E\xc$ with basis sets has also led to an obvious flaw in the ALDA kernel at short distances\cite{FV05}.\\
	
Further work is needed to find accurate XC kernels. One method\cite{LGP00} to test these is by examining the uniform electron gas as the frequency dependend susceptibility 
can be found easily and uses the well known Lindhard function. Hence different approximate XC kernels may be tested and their results compared to highly accurate 
Monte-carlo simulations.

\subsection{Strong fields}
\label{s:strong}

Next we turn our attention to the non-perturbative regime.  Due to advances
in laser technology over the past decade, many experiments are
now possible in regimes where the laser field is stronger than the
nuclear attraction\cite{MG04}. The time-dependent field cannot
be treated perturbatively, and even solving the time-dependent
Schr\"odinger equation in three dimensions for the evolution of
two interacting electrons is barely feasible with 
present-day computer technology\cite{PMDT00}.\\

For more electrons in a time-dependent field,
wavefunction methods are 
prohibitive, and in the regime of (not too high) laser intensities, 
where the electron-electron interaction is
still of importance, TDDFT is essentially the only practical scheme
available\cite{KLG04,KGLG03,PG99,LGE02,LKGE02,LEG01,LGE00}.  There are a whole host of phenomena that TDDFT
might be able to predict:  high harmonic generation, multi-photon
ionization, above-threshold ionization, above-threshold dissociation, etc., but 
only if accurate approximations are available.

%

With the recent advent of
atto-second laser pulses, the electronic time-scale has become accessible. 
Theoretical tools to analyze the dynamics of excitation processes on
the attosecond time scale will become more and more important.
An example of such a tool is the time-dependent electron
localization function (TDELF) \cite{BMG04,EGV04}. This quantity
allows the time-resolved observation of the formation, 
modulation, and breaking of chemical bonds, 
thus providing a visual understanding of the dynamics of excited electrons 
(for an example see 
Ref. \cite{WG04}).
The natural way of calculating the TDELF is from the TDKS orbitals. \\

High harmonic generation (HHG) is the production from medium intensity
lasers of very many harmonics (sometimes hundreds) of the input
intensity.
Here TDDFT calculations
have been rather succesful for atoms \cite{UEG96,EG99,CMAB04} and
molecules\cite{CC01,BNZM03}.  Recent experiments have used
the HHG response of molecules to determine their vibrational
modes\cite{WWCP06}.  Calculations have been performed using
traditional scattering theory\cite{Greene}.  If this method
grows to be a new spectroscopy, perhaps the electron scattering
theory of Sec \ref{s:scatt} will be used to treat large molecules.

Multi-photon ionization occurs when an atom or molecule loses
more than one electron in an intense field.
About a decade ago, this was discovered to be a non-sequential
process, i.e., the probability of double ionization can be much
greater than the product of two independent ionization events,
leading to a 'knee' in the double ionization probability as a function
of intensity\cite{WSMA94,FBCK92,KSTN93}.  TDDFT calculations have so far been unable
to accurately reproduce this knee, and it has recently been shown
that a correlation-induced derivative discontinuity is needed
in the time-dependent KS potential\cite{LK05}.

Above-threshold ionization (ATI) refers to the probability of ionization
when the laser frequency is less than the ionization potential, i.e.,
it does not occur in linear response\cite{PRS00,NBU04}.  
Again, this is not well given by TDDFT calculations, but both this
and MPI require knowledge of the correlated wavefunction, which is
not directly available in a KS calculation.

Since the ionization threshold plays a crucial role in most strong
field phenomena, 
Koopmans theorem relating the energy level of the KS HOMO to
the ionization energy must be well satisfied. This suggests the use
of self-interaction free methods such as OEP\cite{GKKG98,UGG95} or LDA-SIC
rather than the usual DFT approximations (LDA,GGA,etc), with their
poor potentials (see Fig. \ref{f:LDA_vs_exKS} in Sec. \ref{s:gsappr}).

The field of quantum control has mainly concentrated on 
the motion of the nuclear wave packet on a given set of
precalculated potential energy surfaces, the ultimate goal being the 
femto-second control of chemical reactions \cite{RZ00}. 
With the advent of atto-second pulses, control of electronic dynamics has
 come within reach. A marriage
of optimal-control theory with TDDFT appears to be the ideal 
theoretical tool to tackle these problems\cite{WG05,UGSR04}.
Recent work\cite{WU04,UV01,UV00} has shown the ability of
TDDFT to predict the coherent control of quantum wells using
Terahertz lasers.
However they remains many difficulties and challenges, including the 
coupling between nuclei and electrons\cite{BN06,LB06,BKB04}, in order to develop a general purpose theory.

\subsection{Transport}
\label{s:trans}

There is enormous interest in transport through single molecules as a key component
in future nanotechnology\cite{NR03}.  Present formulations use ground-state
density functionals to describe the stationary non-equilibrium
current-carrying state\cite{BMOT02}.  But several recent suggestions consider
this as a time-dependent problem\cite{SA04,VT04,GCb04,BCG05},
and use TD(C)DFT for a full description
of the situation.  Only time will tell if TDDFT is really needed for 
an accurate description of these devices.


\begin{figure}[tbh]
\begin{center}
\leavevmode
\includegraphics[angle=0,width=7cm]{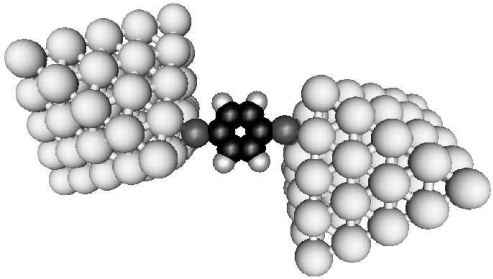}
\end{center}
\caption{Schematic representation of a
  benzene-1,4-di-thiol molecule between two gold contacts. The
  molecule plus gold pyramids (55 atoms each) constitute the {\em extended
  molecule} as used in the DFT calculations for the Landauer approach.}
\label{f:benz}
\end{figure}

Imagine the setup shown in Fig. \ref{f:benz} where a conducting molecule is sandwiched between two contacts which are connected to semi-infinity leads. The Landauer formula for the current is
\ben
\label{eq:land}
I = \frac{1}{\pi}\int_{-\infty}^{\infty}dE~ T(E)[f_{L}(E)-f_{R}(E)]
\een
where $T(E)$ is the transmission probability for a given energy and $f_{L/R}(E)$ is the Fermi distribution function for the left/right lead. The transmission probability can be written using the non-equilibrium Green's functions (NEGF) of the system. Ground-state DFT is used to find the KS orbitals and energies of the extended molecule and used to find the self-energies of the leads. These are then fed into the NEGF method, which will determine $T(E)$ and hence the current.\\

The NEGF scheme has had a number of successes, most notably for atomic-scale point contacts and metallic wires. Generally it does well for systems where the conductance it high. 
However, it was found that for molecular wires, the conductance is overestimated by $1-3$ orders of magnitude. Various explanations for this and the problems with DFT 
combined with NEGF in general have been suggested.\\

Firstly, the use of the KS orbitals and energy levels has no theoretical basis. The KS orbitals are those orbitals for the non-interacting problem that reproduce the correct ground-state density. They should not be thought of as the true single-particle excitations of the true system. However as we have seen they 
often reproduce these excitations qualitatively, so it is not clear 
to what extent this problem affects the conductance.\\

The geometry of the molecules was also suggested as a source of error. DFT first relaxes the molecule to find its geometry, whereas in the experiments the molecule
may be subject to various stresses that could rotate parts of it and/or squash parts together. However calculations have shown that the geometry corrections are small\cite{EWK04}.\\

The approximation that the non-equilibrium XC functional is the same as for the static case has been suggested as a major source of error. In fact neither the HK theorem 
nor the RG theorem are strictly valid for current-carrying systems in homogeneous electric fields.  A dynamical correction to the LDA functional for the static case has been derived using the Vignale-Kohn functional TDCDFT but were found to yield only small corrections to ALDA\cite{SZVD05}.\\

In a similar vein, the lack of the derivative discontinuity and self interacting errors (SIE) in the approximations to the XC functional may be the source of the
problem\cite{EWK04}. In Hartree-Fock (HF) calculations (and also in OPM 
calculations\cite{KKP04} with EXX, exact exchange), which have
no SIE, the conductances come out a lot lower in most regions\cite{KBY06}. Also 
calculations have been done using a simple model\cite{TFSB05} with a KS potential with a derivative
discontinuity. The I-V curves for this system are significantly different from those predicted by LDA. This problem is most severe when the molecule is not strongly coupled to the leads, but goes away when it is covalently bonded. Recent OEP calculations of the transmission along a H-atom chain verify these features\cite{KBY06}.\\

Despite these problems, quantitative results can be found for molecular devices. By looking at what bias a KS energy level gets moved between the two chemical potentials
of the leads ( and hence by Eq. (\ref{eq:land}) there should be a conductance peak), one can qualitatively predict postitions of these peaks\cite{KBE06}, although the magnitude of the
conductance may be incorrect by orders of magnitude.\\

Since transport is a non-equilibrium process, we should expect that using static DFT will not be able to accurately predict all the features. Recently a number of methods 
have been suggested to use TDDFT to calculate transport.
In Ref. \cite{KSAR05}, the authors present a practical scheme using TDDFT to calculate current. The basic idea is to 'pump' the system into a non-equilibrium initial state
by some external bias and then allow the KS orbitals to evolve in time via the TDKS equations. The RG theorem then allows one to extract the longitudinal current using
the continuity equation. Using transparent boundary conditions in the leads (these solve problems with propagating KS in the semi-infinite leads) and using an iterative procedure
to get the correct initial state, they are able to find the steady state current.

\begin{figure}
\includegraphics[width=3.3in]{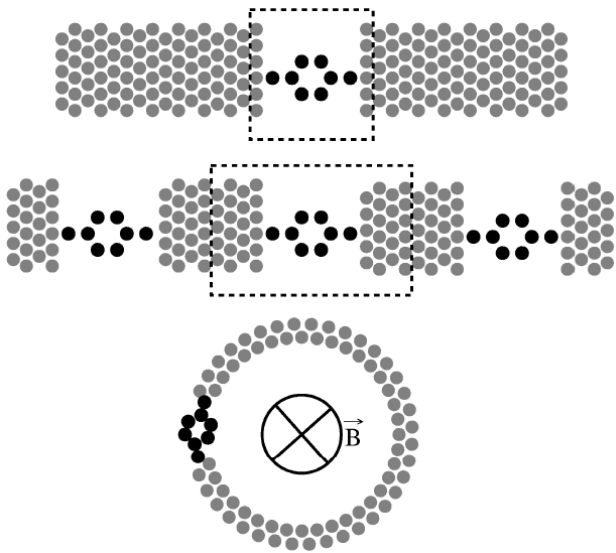}
\caption{\label{f:ring_geo} Ring geometry for gauge transformation of electric fields.}
\end{figure}

An alternative formulation uses periodic boundary conditions and includes dissipation\cite{GBC06}.
In the Landauer-B\"{u}ttiker formulism, dissipation effects due to electron-electron interaction and electron-phonon interaction are neglected as the molecule is smaller than the scattering length. However, there would be scattering in the leads. Imagine a molecule in the ring geometry, with a spatially constant electric field. Via a gauge transformation, this can be replaced by a constant time-dependent magnetic field through the center of the ring. If there is no dissipation, the electrons would keep accelerating indefinitely and never reach a steady state.\\

In the classical Boltzmann equation for transport, scattering is included via a dissipation term using $\tau$, the average collision time. A master equation approach is basically a generalization of the Boltzmann equation to a fully quantum mechanical system.
The master equation is based on the Louville equation in quantum mechanics and for a quantum mechanical density coupled to a heat bath (or reservoir), it is written as
\ben
\frac{d}{dt} \hat{\rho}(t) = -\imath[H,\hat{\rho}(t)] + \mathbf{C}[ \hat{\rho}(t) ]
\een
where $\mathbf{C}$ is a superoperator acting on the density whose elements are calculated using Fermi's Golden rule with $V_{el-ph}$ in a certain approximation (weak coupling and instantaneous processes).
A KS master equation\cite{BCG05} can be set up, modifying $\mathbf{C}$ for single particle reduced density matrices so that it will give the correct steady state. The TDKS equations are then used to propagate forward in time until the correct steady state density is found. The current in then extracted from this.
Recent calculations have shown it can give correct behaviour, such as hysteresis in I-V curves.\cite{GCb04,GCc04}\\
%
%
%
%

\section{Summary}
\label{s:sum}

We hope we have conveyed some of the spirit and excitement of TDDFT in
this non-comprehensive review.   We have explained what TDDFT is, and
where it comes from.  We have shown that it is being used, and often works
well, for many molecular excitations.  Its usefulness lies neither in
high accuracy nor reliability, but in its qualitative ability to
yield roughly correct absorption spectra for molecules of perhaps several
hundred atoms.  Thus we emphasize that, usually, there are many excitations
of the same symmetry, all coupled together, and that these are the circumstances
under which the theory should be tested.  For many molecular systems,
TDDFT is now a routine tool that produces useful accuracy with reasonable
confidence.

That said, we have discussed some of the areas where TDDFT in its current
incarnation is not working, such as double excitations, charge transfer, and
extended systems.  But there has been significant progress in two out of three of
these, both in understanding the origin of the problem, and finding alternative
approaches that may
ultimately yield a practical solution.  We also studied how well TDDFT works
for a few cases where the exact ground-state solution is known, describing the
accuracy of different functionals.  We also surveyed some applications beyond
simple linear response for optical absorption, such as ground-state
functionals from the adiabatic connection, strong fields, and transport.
In each of these areas, more development work seems needed before TDDFT calculations
can become a routine tool with useful accuracy.

Many wonder how long DFT's preemminence in electronic
structure can last.  For sure, Kohn-Sham DFT is a poor player that
struts and frets his hour upon the stage of electronic
structure, and then is heard no more.  After all, its
predecessor, Thomas-Fermi theory, is now obsolete, being
too inaccurate for modern needs. Many alternatives for electronic excitations, such as GW, are becoming computationally
feasible for interesting systems.   But we believe DFT, and TDDFT,
should dominate for a few decades yet.

We thank Michael Vitarelli for early work and Dr. Meta van Faassen and Dr. Max Koentopp for providing figures.
K.B. gratefully acknowledges support of the US Department of Energy, under
grant number DE-FG02-01ER45928, and the NSF, under grant CHE-0355405.
This work was supported, in part, 
by the Center for Functional Nanostructures (CFN) of the Deutsche
Forschungsgemeinschaft (DFG) within project C3.9, the EXC!TiNG
Research and Training 
Network of the European Union and the NANOQUANTA Network of Excellence.

\end{document}